\documentclass[]{aastex61}
\usepackage{multirow}
\usepackage{captcont}
\usepackage{amsmath} 

\received{XXX}
\revised{XXX}
\accepted{XXX}

\shorttitle{Planet formation and disk evolution }
\shortauthors{van der Marel \& Mulders}

\begin{document}

\title{A stellar mass dependence of structured disks: a possible link with exoplanet demographics}

\correspondingauthor{Nienke van der Marel}
\email{astro@nienkevandermarel.com}

\author{Nienke van der Marel}
\affil{Physics \& Astronomy Department, 
University of Victoria, 
3800 Finnerty Road, 
Victoria, BC, V8P 5C2, 
Canada}

\author{Gijs D. Mulders}
\affil{Facultad de Ingenier\'{i}a y Ciencias, Universidad Adolfo Ib\'{a}\~{n}ez, Av.\ Diagonal las Torres 2640, Pe\~{n}alol\'{e}n, Santiago, Chile}
\affil{Millennium Institute for Astrophysics, Chile}

\begin{abstract}
Gaps in protoplanetary disks have long been hailed as signposts of planet formation. However, a direct link between exoplanets and disks remains hard to identify. We present a large sample study of ALMA disk surveys of nearby star-forming regions to disentangle this connection.
All disks are classified as either structured (transition, ring, extended) or non-structured (compact) disks. Although low-resolution observations may not identify large scale substructure, we assume that an extended disk must contain substructure from a dust evolution argument. A comparison across ages reveals that structured disks retain high dust~masses up to at least 10 Myr, whereas the dust~mass of compact, non-structured disks decreases over time. This can be understood if the dust~mass evolves primarily by radial drift, unless drift is prevented by pressure bumps. We identify a stellar~mass dependence of the fraction of structured disks. We propose a scenario linking this dependence with that of giant exoplanet occurrence rates. We show that there are enough exoplanets to account for the observed disk structures if transitional disks are created by exoplanets more massive than Jupiter, and ring disks by exoplanets more massive than Neptune, under the assumption that most of those planets eventually migrate inwards. 
On the other hand, the known anti-correlation between transiting super-Earths and stellar~mass implies those planets must form in the disks without observed structure,
consistent with formation through pebble accretion in drift-dominated disks. These findings support an evolutionary scenario where the early formation of giant planets determines the disk's dust evolution and its observational appearance.
\end{abstract}



\section{Introduction}

The past decade has seen a wealth of new observational data on both exoplanets and protoplanetary disks, allowing direct comparisons and statistical studies to help answer questions about the planet formation process. Demographic surveys of disks in nearby star forming regions with the Atacama Large Millimeter/submillimeter Array (ALMA) have resulted in hundreds of measurements of protoplanetary disk dust masses \citep[e.g.][]{Ansdell2016,Ansdell2018,Pascucci2016,Barenfeld2016,Cieza2019}, which, together with studies of stellar properties such as accretion rate and stellar mass, have resulted in new insights on the disk evolution process. For example, disk dust masses correlate with stellar mass ($M_{\rm dust}-M_*$ relation), but this correlation steepens with time, with a stronger drop for low-mass stars \citep{Pascucci2016,Ansdell2017}. The disk dust radius is correlated with the disk dust mass \citep[the 'size-luminosity' relation,][]{Andrews2013,Tripathi2017} and the quadratic dependence is generally consistent with radial drift dominated disks, albeit with large scatter \citep{Rosotti2019}. Efficient inward radial drift is further supported by a significant decrease of disk dust size with age across regions \citep{Hendler2020}. On the other hand, pressure bumps have been proposed to explain the existence of massive dust disks, preventing radial drift \citep{Pinilla2012a,Pinilla2020}, which is also suggested to explain the scatter in the size-luminosity relation \citep{Rosotti2019}. Finally, protoplanetary disk dust masses appear low compared to the solid content of observed exoplanets, suggesting early planet formation \citep{Manara2018,Tychoniec2020}, though a careful accounting of exoplanet detection biases reveals that disks contain similar amounts of solids as found in exoplanetary systems (\citealt{NajitaKenyon2014}, Mulders et al., subm. to ApJ).

Meanwhile, our understanding of true exoplanet demographics has improved significantly, despite the inherent biases in the detection techniques. With proper bias corrections, trends in planet occurrence rates have been estimated for a wide range of stellar and planetary properties. 
Generally, the most massive planets are also found to be the rarest, with planets becoming increasingly more common towards lower masses or smaller sizes \citep[e.g.][]{Mayor2011,Howard2010,Howard2012, Suzuki2016,Wagner2019}. Gas giant planets, here loosely defined as planets more massive than Neptune, orbit 10-20\% of stars \citep{Cumming2008,Fernandes2019}. Furthermore, hot Jupiters are rare ($<1\%$) while giant planets become increasingly more common at larger separations from the star \citep{Cumming2008,DongZhu2013,Santerne2016}. 
The giant planet occurrence rate then peaks at 2-3 au \citep{Fernandes2019}, and decreases again in the regions accessible to direct imaging \citep{Nielsen2019,Wagner2019}. 
The occurrence rate of Jupiter-mass planets is positively correlated with stellar mass \citep{Johnson2010,Reffert2015}. Planets smaller in size than Neptune show a different dependence on stellar properties. These planets --- studied in detail with the Kepler transit survey and often referred to as super-Earths, mini-Neptunes, or sub-Neptunes --- orbit nearly half of all sun-like stars \citep{Zhu2018,Mulders2018,He2019}. These sub-Neptunes are more common around M-dwarfs than Sun-like stars \citep{Mulders2015a,Mulders2018} and their occurrence also varies between stars with spectral type G, K, and F \citep{Howard2012,Hsu2019,Yang2020,He2020}. 

The most direct link between protoplanetary disks and planets is the discovery of gapped protoplanetary disks or ring disks, with one or more narrow gaps in their dust distribution, such as HL Tau and HD~163296 \citep{HLTau2015,Isella2016}. Dust gaps have been proposed to be caused by clearing by a companion \citep{LinPapaloizou1979,Pinilla2012b}, where the dust gets trapped in a pressure bump just outside the planet gap, prevented from drifting inwards \citep{Weidenschilling1977,Birnstiel2010}. Other explanations for dust gaps have been suggested as well, such as snow lines \citep{Zhang2015}, zonal flows \citep{Johansen2009} and secular gravitational instabilities \citep{Takahashi2016}. The snow line scenario is unable to explain the majority of ring locations \citep{Huang2018,Long2018,vanderMarel2019}, implying that this cannot be the main mechanism responsible for dust gaps.

The planet clearing scenario followed by dust trapping is supported by observations of deep gas gaps well inside the dust gaps \citep[e.g.][]{vanderMarel2016}, in so-called transition disks (TDs): disks with a large ($\gtrsim$20 au) cleared inner dust cavity. Transition disks were originally identified in the infrared in their Spectral Energy Distribution \citep{Strom1989,Espaillat2014}, and dozens of disks with large dust cavities of tens of au have been revealed with millimeter interferometry by both pre-ALMA and ALMA programs \citep[e.g.][]{Andrews2011,Francis2020}. Gas gaps in transition disks have been linked to Super-Jovian planets on eccentric orbits \citep{Muley2019,vanderMarel2020} or even stellar binaries \citep{Price2018}. Planetary companions have indeed been found in the PDS~70 disk cavity \citep{Keppler2018,Haffert2019}, but for many disk cavities, the limits from direct imaging close to the star are not stringent enough to confirm their origin \citep[see][for an overview]{vanderMarel2020}. 

For ring disks, gas gaps are harder to characterize \citep{vanderMarel2018-hd16,Rab2019}, but kinematic signatures hint at the presence of companions in their gaps  \citep{Pinte2018,Pinte2019,Pinte2020,Teague2018,Teague2019}. Quantitatively,  gaps in ring disks have been linked to (moderately) massive giant planets based on their observed gap width, ranging from about a Neptune mass (0.05 $M_{\rm Jup}$) to $\sim$few $M_{\rm Jup}$ \citep{Zhang2018,Lodato2019}. This can be understood by the theoretical minimum required planet mass to create a pressure bump at large radii that limits radial drift, which is estimated at $\sim$10-20 $M_{\rm Earth}$ or 0.6-1.2 $M_{\rm Nep}$ depending on disk and stellar properties \citep{Rosotti2016,Sinclair2020}. In summary, planets clearing gaps and trapping dust outside their orbits appear to be a plausible scenario to explain the observed dust gap morphologies.

Matching disk demographics with exoplanet demographics of main sequence stars remains challenging: the occurrence rates of giant exoplanets on wide orbits similar to gap radii appear too low to be comparable with the fraction of transition disks of $\sim$11\% \citep{vanderMarel2018}. Interestingly, transition disks have been found to be more common in massive disks and around more massive stars \citep{Owen2012}, which trend is consistent with giant exoplanets on wide orbits \citep{Nielsen2019}. A larger number of Jovian and sub-Jovian planets has been identified in radial velocity surveys around sun-like stars, but those planets are typically located closer to the star at a few au \citep{Johnson2010, Fernandes2019}. However, it is possible that planets migrate inwards during the lifetime of the disk \citep{KleyNelson2012} and the gap location is less relevant for the comparison with exoplanets \citep{Lodato2019}.

The fraction of ring disks in the total disk population remains unknown. Although ring disks appear to be common in high resolution ALMA observations, the high-resolution observations so far are heavily biased towards the brightest disks, as these are easier to observe \citep{Andrews2018}. The Taurus disk survey by \citet{Long2019} shows that structured disks (transition and ring disks) are indeed not that common for a relatively unbiased sample selection, but statistical conclusions remain challenging due to their exclusion of previously observed disks. Interestingly, structured disks appear to maintain their millimeter dust longer than most disks: in several older regions where most disk dust masses are low ($<1 M_{\rm Earth}$), a handful of massive outlier disks have been identified and their presence has been linked to pressure bumps which limit radial drift \citep{Ansdell2020,Michel2020}. 

Overall there is no good understanding of the commonality of transition disks, ring disks and apparent non-structured disks and their role in the dust evolution processes such as radial drift. As structured disks, or at least massive disks, are still rare, disk surveys are dominated by low-mass non-structured disks, which can explain why dust evolution models including radial drift generally hold. On the other hand, structured disks are often studied in much more detail and higher resolution due to their more apparent link with planets. The only high-resolution observations of a compact dust disk is that of CX Tau \citep{Facchini2019}, which is no more than 14 au in radius and does not show substructure at a spatial resolution of 40 mas ($\sim$5 au).

In this work we construct a sample of dust disks from all nearby star forming regions studied by ALMA, and analyze their morphology. Despite the range of spatial resolutions and sensitivities we aim to use any information that can be reasonably deduced from the datasets to classify their morphology in the millimeter dust continuum,  in particular the presence of large scale gaps and radial dust size. We make the explicit assumption here that the dust size and morphology is entirely regulated by radial drift and trapping in pressure bumps, following \citet{Pinilla2020}. We look for trends with their age and stellar properties and quantify the fraction of disks with observed large scale substructure for different stellar mass bins. We then make a comparison with exoplanet demographics of main-sequence stars, with the assumption that all dust structures are caused by planets. The aim of this comparison is to test whether enough planets are available to cause the observed disk structures, and what the minimum mass of a planet to create a transition disk or dust ring would have to be.

This paper is organized as follows. Section \ref{sct:sample} describes the sample selection and the origin of the data. In Section \ref{sct:analysis} we describe our choices in the disk morphology classification and Section \ref{sct:results} presents a number of plots showing the trends with age and stellar mass for disks. In Section \ref{sct:exoplanets} we construct a demographical model of exoplanets and compare the stellar mass dependence with that of protoplanetary disk structures. In Section \ref{sct:discussion} we describe the implications of our study for our understanding of disk morphology and disk evolution.

\section{Sample}
\label{sct:sample}
\subsection{Sample section}
Our sample consists of millimeter continuum fluxes from all known protoplanetary disk surveys from the literature: Ophiuchus \citep{Cieza2019}, Taurus \citep[][]{Akeson2019}, Chamaeleon \citep[][]{Pascucci2016}, Lupus  \citep[][]{Ansdell2018}, CrA \citep{Cazzoletti2019}, IC348 \citep{Ruiz-Rodriguez2018}, Upper Sco \citep{Barenfeld2016} and TW Hya \citep{Rodriguez2015}, and fluxes of disks in $\epsilon$ Cha and $\eta$ Cha from Aguayo et al. in prep. The main references for the disk surveys of each region are summarized in Table \ref{tbl:regions}. 

For most regions, we take the entire selection from the corresponding ALMA disk survey, which generally encompasses a complete sample of Class II objects as identified from previous \textit{Spitzer} studies. Class II objects are generally considered as protoplanetary disks according to the Lada classification \citep{Lada1984}. Targets with later spectral type than M6 (stellar mass $\leq$0.08 $M_\odot$) are excluded as the disk surveys are likely incomplete for these low-mass stars and a statistical comparison becomes unreliable. For Ophiuchus, Upper Sco, $\epsilon$ Cha and $\eta$ Cha and Taurus the sample selection was a bit more complicated due to the nature of the disk surveys as described below.

\paragraph{Ophiuchus}
For Ophiuchus, Class I and F targets are included in our sample following the same approach as  the corresponding ALMA survey \citep{Cieza2019}. The reason is that Ophiuchus suffers from high extinction by foreground clouds \citep[e.g.][]{McClure2010}, which can make protoplanetary disks appear as Class I or Flat objects. Infrared color criteria such as the Lada classification are not a good indicator of evolutionary stage in that case. Excluding Class I and Flat sources from the survey would remove a large number of protoplanetary disks.

\paragraph{UpperSco}
The Upper Sco disk survey by \citet{Barenfeld2016} was selected using different color criteria than the Lada criteria used by most of the other ALMA disk surveys, and it turns out that this sample contains a large number of Class III objects \citep{Michel2020}. In this work, we only consider the Class II objects from Barenfeld's survey for consistency with the other disk surveys. 

\paragraph{$\epsilon$ Cha and $\eta$ Cha}
The samples of $\epsilon$ Cha and $\eta$ Cha are semi-complete using currently available ALMA observations, with 8 of the 12 Class II objects in $\epsilon$ Cha as identified in \citet{Murphy2013} and 3 of the 6 Class II objects in $\eta$ Cha as identified in \citet{Sicilia2009}. 

\paragraph{Taurus}
For Taurus, we adopt the target list by \citet{Akeson2019}, who constructed a full millimeter continuum sample of Class II objects based on e.g. \citet{Andrews2013,Akeson2014,WardDuong2018} and their own work. All but \citet{Andrews2013} provide moderate resolution ALMA observations (0.2-0.4"), covering about 65\% of their sample. The remaining 82 targets with millimeter data are based on low resolution SMA fluxes from \citet{Andrews2013}. However, 33 of these have been reobserved with ALMA at higher resolution in e.g. \citet[][and a few individual works, see Table \ref{tbl:fullsample}]{Long2019,Francis2020} and 8 disks have been analyzed with visibility modeling by \citet{Tripathi2017}, so we use these constraints. For the remaining SMA targets, 15 upper limits (above 3 mJy) are removed as they do not constrain at a similar level as the ALMA disk surveys. The remaining 26 targets do not have any information about their structure or size, but most of their fluxes are relatively low (up to 30 mJy). Although it is possible that some of these are structured or extended, the fraction is very small compared to the total number of Taurus targets, so they are included. Finally, we add transition disks AB~Aur, MWC~758, CQ~Tau and GM~Aur \citep{Francis2020} and MWC~480, CI~Tau and T~Tau from \citet{Long2019} to the sample, which were not included by \citet{Akeson2019}. The final sample consists of 167 targets in Taurus. 

All targets have been compared with \textit{Gaia DR2}. Targets that can be excluded as non-members based on their \textit{Gaia} distance are excluded from our sample. Stellar properties are taken from the literature (see Section \ref{sct:stellar} for details). 

The total sample consists of 700 disks. Table \ref{tbl:regions} provides an overview per region, including distance, age and references for the disk surveys and stellar properties. The individual targets and their properties are listed in Table \ref{tbl:fullsample}.

\subsection{Spatial resolution}
The spatial resolution is relevant for the detectability of large scale substructures such as gaps and rings at scales of tens of au, or in absence of detectable substructure, to determine the disk dust size. Disk surveys have been performed at a range of spatial resolutions and are located at various distances. Table \ref{tbl:regions} lists the approximate spatial resolution for each region.  

\begin{table}[!ht]
\centering
\caption{Overview regions}
\label{tbl:regions}
\begin{tabular}{lllllll}
\hline
Region&$<d>$&Age&$N$&Resolution&Resolution&Refs$^a$\\
&(pc)&(Myr)&&(")&(au)&\\
\hline
Ophiuchus&140&1-2&163&0.2&28&1,2\\
Taurus&145&1-2&167&$\sim$0.2-0.4&$\sim$30-60&3,4\\ 
Lupus&160&1-3&94&0.25&40&5,6,7\\
Cham I&180&1-6&85&0.6$^b$&108&8,9,7\\
CrA&150&1-5&39&0.32&39&10,11\\ 
IC348&250&2-3&59&0.8$^b$&200&12,13\\ 
$\epsilon$ Cha&110&3-8&9&0.25&28&14,15\\
TW Hya&56&7-13&4&$\sim$1.0&$\sim$60&16,17,18\\ 
$\eta$ Cha&94&8-14&2&0.25&28&16,15,19\\
Upper Sco&145&10-12&72&0.34&49&20,21,22\\ 
Lower Cen Crux (LCC)&118&12-18&2&$\sim$0.2&$\sim$24&23,24\\ 
Upper Cen Lupus (UCL)&140&14-18&4&$\sim$0.2&$\sim$28&23,24\\ 
\hline
\end{tabular}\\
$^a$ References for age, disk survey and stellar properties, respectively.
1) \citet{Wilking2008}; 2) \citet{Cieza2019}; 3) \citet{Kraus2009}; 4) \citet{Akeson2019}; 5) \citet{Comeron2008}; 6) \citet{Ansdell2018}; 7) \citet{Manara2018}; 8) \citet{Luhman2007}; 9)\citet{Pascucci2016}; 10) \citet{Meyer2009}; 11) \citet{Cazzoletti2019}; 12) \citet{Luhman2003}; 13) \citet{RuizRodriguez2018}; 14) \citet{Murphy2013}; 15) Aguayo et al. in prep.; 16) \citet{Bell2015}; 17) \citet{Rodriguez2015}; 18) \citet{Venuti2019}; 19) \citet{Rugel2018}; 20)  \citet{Esplin2018}; 21) \citet{Barenfeld2016}; 22) \citet{Manara2020}; 23) \citet{Pecaut2016}; 24) Individual studies, see Table \ref{tbl:fullsample}.\\
$^b$ Although the beam size of the images is relatively large for Chamaeleon and IC348, the performed \textit{uvmodelfit} fitting by the authors results in recovered diameters as small as 0.11"/20 au (Chamaeleon) and 0.21"/52 au (IC348) comparable to the resolution of other surveys.
\end{table}

Most disk surveys have a spatial resolution of $\sim$50 au (25 au radius) or less, except for IC348 and Chamaeleon, which have been observed at lower spatial resolution at $\sim$0.6-0.8" ($>$100 au) \citep{Pascucci2016,RuizRodriguez2018}. However, both papers model the disk size in the \textit{uv}-plane using the CASA task \texttt{uvmodelfit}, which provides size estimates down to 0.2" diameter ($\sim$40 au), which is comparable to the spatial resolution of other surveys and thus sufficient for our purposes. 

\subsection{On the UCL/LCC disks}
Protoplanetary disks in the older Upper Centaurus Lupus and Lower Centaurus Crux (UCL/LCC) regions are not studied systematically in millimetre wavelengths, as the stellar population is thought to be much older than the other regions in our study, with a median age of $\sim$15 Myr \citep{Pecaut2016}. A handful of exceptions are the famous, bright transition disks HD142527, HD135344B, PDS70 and AK~Sco in UCL, and HD100546 and HD100453 in LCC. Interestingly, the latter two are the only massive disks identified in an infrared study of A-stars in Sco-Cen by \citet{Chen2012}, with the third massive disk having an estimated dust mass of only 1.5 $M_{\rm lunar}$. Also AK~Sco, HD139614 and MP~Mus have been identified as primordial disks in UCL and LCC \citep{Chen2011,Preibisch2008}, with massive infrared disks in contrast with the bulk of the disk populations which have disk masses $<$2 $M_{\rm lunar}$. For MP~Mus and HD139614 no high resolution data are available, so it is unclear whether they are structured and marked as such. 

In addition, the isolated structured disks HD~163296 \citep{Isella2016} and HD~169142 \citep{Fedele2017} are included in our sample. These disks are located just south-east of the Upper Sco region and have individual ages of 12 $\pm$3 and 7$\pm$2 Myr, respectively \citep{vanderMarel2019}. Millimetre dust masses are derived from available ALMA fluxes (see Table \ref{tbl:fullsample}). No other millimetre fluxes of UCL/LCC members have been derived to our knowledge, but considering the infrared studies mentioned above it is unlikely that any other disks are in the $>0.1 M_{\rm Earth}$ regime. The UCL/LCC and isolated disks are thus only included in the age comparison in Figure \ref{fig:massdistribution} to demonstrate the survival of massive disks, but not in the stellar mass diagrams to avoid inclusion of highly incomplete samples of these older regions.

\section{Analysis}
\label{sct:analysis}

\subsection{Disk classification}
\label{sct:classification}
All disks are initially classified as either transition disk, ring disk or 'non-structured' (no large scale structure identified). Large scale structure is defined as the presence of gaps at scales of 25 au or more, i.e. with a gap edge located at at least 25 au radius. The classification is based entirely on the literature classification. We have thus not reanalyzed the disk images ourselves, as gaps are often only revealed through visibility modeling. This means that transition and ring disks are classified as such as identified by the disk survey papers from Table \ref{tbl:regions}, but we have also compared our list with transition disk surveys which have derived cavity sizes \citep{vanderMarel2018,Pinilla2018,Francis2020} and ring disk studies \citep{Andrews2018,Long2019} for our classification. The distinction between a transition disk and ring disk is as follows: a ring disk is defined as a disk with one or multiple dust gaps throughout the disk, but with dust emission in the center, whereas a transition disk is defined as a disk with an inner cleared dust gap of at least 25 au in radius, without significant dust emission at the central location. The threshold of 25 au is chosen based on the largest gap widths seen in ring disks \citep{Huang2018}. Transition disks are only classified as such when the millimeter continuum image has confirmed the inner cavity, in contrast to classical SED-selected transition disks which are known to be less well defined and often inconsistent in cavity size  \citep[e.g.][]{Brown2009,vanderMarel2018}. Transition disks with additional gaps in the outer disk are still classified as transition disk. 

With this classification of large scale structure, the total sample contains 42 transition disks and 37 ring disks out of 700 disks. When the incomplete regions UCL, LCC and isolated sources are excluded, as there is no information about the non-structured disks in their corresponding region, these numbers become 37 transition disks and 30 ring disks out of 692 disks: about 10\% of the sample is thus structured for the given resolution. 

This classification obviously underestimates the number of structured disks, as the majority of the disks have not been observed at sufficiently high angular resolution to detect large scale gaps. However, dust evolution models have shown that pressure bumps are required to explain the presence of millimeter dust grains at large orbital radii to limit the radial drift of millimeter grains \citep[e.g.][]{Weidenschilling1977, Birnstiel2010,Pinilla2012a}. In the absence of pressure bumps, radial drift would inevitably lead to compact, low-mass dust disks in a few Myr. This implies that large, extended dust disks can only exist when supported by pressure bumps, which could reveal itself as continuum large scale substructure at high resolution. Therefore, the outer disk dust size can provide an alternative way to infer the likely presence of substructure in dust disks, which will be used in our subsequent classification of disks without observed structure. We make the explicit assumption here that all dust substructure is caused by pressure bumps and dust size is entirely regulated by dust evolution and radial drift.

For the `non-structured disks' in our sample, we thus use the dust radius for further classification from a dust evolution argument. The dust radius of each disk $R_{\rm size}$ is estimated as half of the major axis or FWHM. For most disk studies in our sample, the FWHM of each disk is estimated using the CASA task \texttt{imfit} or \texttt{uvmodelfit}. For some studies, the effective radius $R_{\rm eff,68}$ (the radius enclosing 68\% of the total flux) is derived \citep{Tripathi2017,Long2019}, which is comparable to half of the FWHM. The $R_{\rm eff,68}$ radii of the resolved disks in Lupus were recomputed, as \citet{Ansdell2018} had only derived $R_{\rm eff,90}$ values. The dust disk outer radii are reported in Table \ref{tbl:fullsample}. Although this is a rather simple approach, it is sufficient for the purpose of identifying extended and compact disks across this large sample for the purpose of our classification. For 148 targets, no flux was detected and only an upper limit on the flux was derived: there is no information on $R_{\rm size}$, but under the assumption of the size-luminosity relation \citep{Tripathi2017} they can be considered compact. 

In order to estimate a threshold for extended disks, we take a look at the distribution of structured disks and non-structured disks in dust size and dust mass. Dust masses are recomputed with the new Gaia DR2 distances using the commonly used relations in \citet{Ansdell2016,Ansdell2018} following the assumptions of isothermal, optically thin emission of 20 K as demonstrated by \citet{Hildebrand1983}:
\begin{equation}
    M_{\text{dust }}=\frac{F_{\nu} d^{2}}{\kappa_{\nu} B_{\nu}\left(T_{\text{dust}}\right)}
\end{equation}
where $B_{\nu}$ is the Planck function for a characteristic dust temperature, $T_{\text{dust}}$, $\kappa_{\nu}$ the dust grain opacity, $d$ the distance to the target in parsecs, and $F_{\nu}$ the millimeter flux, using the \textit{Gaia} distances when available. Although the dust opacity may change as function of disk age due to grain growth \citep{Testi2014}, we have chosen to follow the general assumptions in the literature of constant dust opacity across age for the purpose of this study.

\begin{figure}[!ht]
    \centering
    \includegraphics{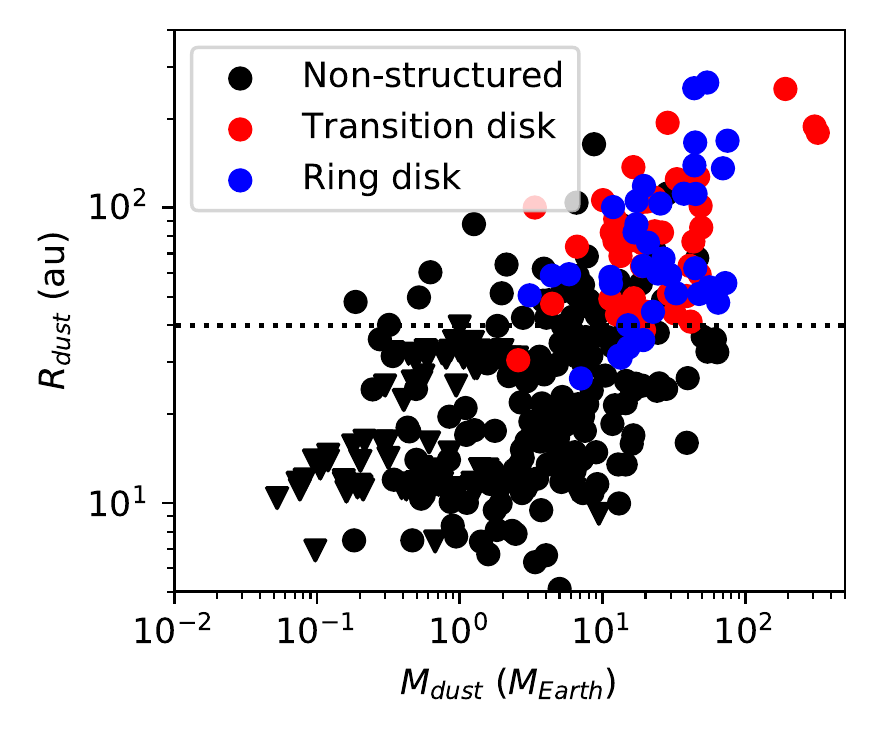}
    \caption{Relation between disk dust size and disk dust mass of detected disks in our sample. Transition disks are marked in red, and ring disks in blue. Disks without known structure are marked black. Upper limits are indicated as triangles, data points as circles. The dotted line indicates the size threshold used in this work to define 'extended' disks (40 au).}
    \label{fig:disksize}
\end{figure}

Figure \ref{fig:disksize} shows the disk size $R_{\rm size}$ ($\sim R_{\rm eff,68}$) versus disk dust mass for all detected dust disks in our sample. This plot is equivalent to the size-luminosity relation \citep{Andrews2010,Tripathi2017,Hendler2020}, showing that lower continuum fluxes generally correspond to smaller dust disks. Although the sizes for the non-structured disks are not derived uniformly, the plot immediately demonstrates that transition and ring disks are large and massive compared to the bulk of the disk population. Transition disks have been known to be at the high end of the disk mass distribution \citep{Owen2012,vanderMarel2018} and also ring disks are generally found in the largest, most massive disks \citep{Andrews2018}, as expected from the presence of pressure bumps \citep{Pinilla2020}. Interestingly, also \citet{Long2019} found a clear distinction in high resolution observations between compact disks without substructure around single stars with $R_{\rm eff,68}$ radii $\lesssim$40 au and disks with substructures with $R_{\rm eff,68}$ radii $\gtrsim$40 au for an unbiased disk survey in Taurus. This motivates the use of a 40 au threshold for the presence of large scale substructure.

Although obviously detection biases exist (see Discussion), the disk dust size provides an approximate tool to identify disks with large scale substructure when the resolution is too low to identify gaps in the dust distribution. We set the threshold above which disks are considered 'Extended' (due to substructure) at 40 au, and include this in our disk classification. In our sample of non-structured disks, we classify 40 targets or $\sim$6\% as 'Extended'. All non-structured disks with sizes $<$40 au are classified as 'Compact'.

\subsection{Stellar properties}
\label{sct:stellar}
\begin{figure}[!ht]
    \centering
    \includegraphics{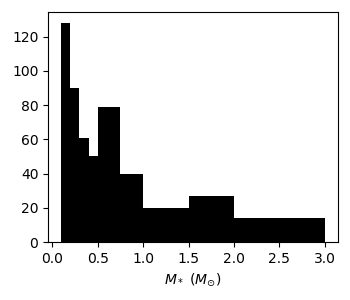}
    \caption{Stellar mass distribution of entire sample. 
    }
    \label{fig:stellarmass}
\end{figure}

Stellar masses have been derived using \textit{Gaia}-corrected luminosities and isochrones for the bulk of our sample; only for Ophiuchus stellar masses remain unconstrained for 93\% of that survey due to the high optical extinction, which prevents proper determination of the stellar properties. For CrA and IC348 the stellar masses have been derived using an average distance, as individual \textit{Gaia} distances were limited \citep{Ruiz-Rodriguez2018,Cazzoletti2019}. As the initial sample selection excluded spectral types outside the B9-M6 range, the final stellar mass distribution ranges from 0.09 to 3 $M_{\odot}$ and is shown in Figure \ref{fig:stellarmass}. This Figure, showing all disks for which the stellar mass has been constrained, represents 73\% of our entire sample. The sample is dominated by subsolar masses, as expected from the stellar mass distributions in star forming regions, but there is also a relatively large amount of intermediate mass stars, due to the additions of bright transition Herbig disks. The sample is thus not fully representative for the IMF, but contains a broad range of targets. 

Age ranges for each region have been derived from evolutionary diagrams of the full population of each region (see references in Table \ref{tbl:regions}). The disk sample is grouped into three categories: young (1-3 Myr), intermediate (3-5 Myr) and old age ($\sim$8-15 Myr) using these values, where the group was based on the mean age of each association. Due to the uncertain nature of the age of CrA \citep[see discussion in][]{Cazzoletti2019} and Chamaeleon, they are included in the intermediate age regime, All regions are indicated in Table \ref{tbl:fit}. LCC and UCL have estimated median ages of $\sim$15 Myr, but the individual age estimates of the 8 disks in our sample are systematically lower \citep[4-12 Myr,][]{Garufi2018}, and likely located in the younger, $\sim$10 Myr regimes of the UCL/LCC complex \citep[][Figure 9]{Pecaut2016}. These disks are thus included in the old age category.

In this work we compare the different disk morphologies as function of age and stellar mass to learn more about the disk evolution of structured and non-structured disks and comparison with exoplanet demographics. The results are presented in Figure \ref{fig:mdiskmstar} to \ref{fig:mstarplanetratios}.

\section{Results}
\label{sct:results}

\begin{figure}
    \centering
    \includegraphics[width=\textwidth]{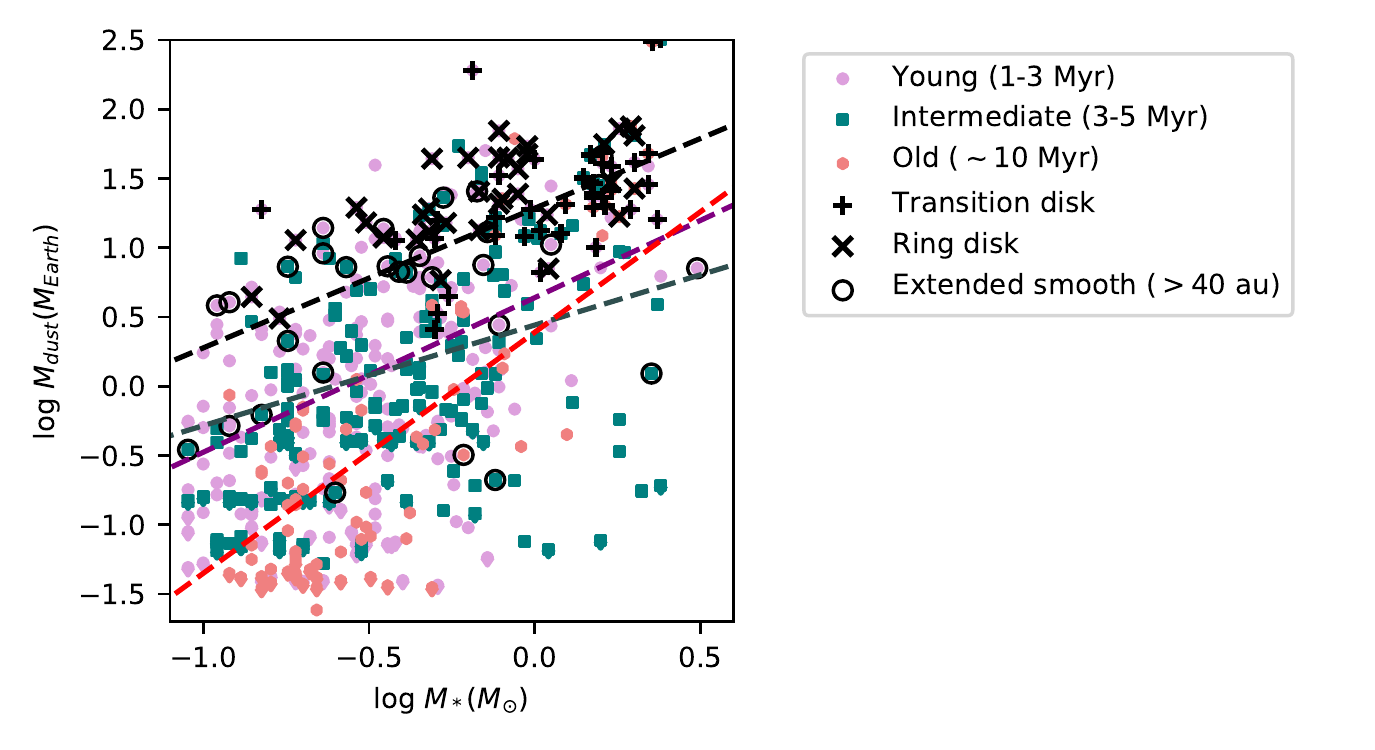}
    \caption{$M_{\rm dust}$-$M_*$ relation, grouped in young (purple), intermediate (teal) and old (coral) regions, as defined in the text. Upper limits are indicated as droplets, data points as circles. The structured disks are marked as transition disk ('+'), ring disk ('x') and extended disk ('o'). The dashed lines show the linear region through each subsample, where the structured disks are not included in the fit of each age-group. The trend is consistent with a faster drop for lower stellar masses.}
    \label{fig:mdiskmstar}
\end{figure}

\begin{table}[!ht]
    \centering
    \caption{$M_{\rm dust}-M_*$ regression fit parameters}
    \label{tbl:fit}
    \begin{tabular}{lllll}
        \hline
         Group  & $\alpha$ (intercept) & $\beta$ (slope) & $\delta$ (scatter)& Regions\\
         \hline
         Young (1-3 Myr) & 0.6$\pm$0.1 & 1.1$\pm$0.2 &0.3$\pm$0.04&Oph,Tau,Lup \\
         Intermediate (3-5 Myr)&0.4$\pm$0.1& 0.7$\pm$0.2 & 0.4$\pm$0.05&Cham,IC348,CrA,$\epsilon$Cha\\
         Old ($\sim$10 Myr) & 0.5$\pm$0.2 & 1.9$\pm$0.3 & 0.3$\pm$0.07&UppSco,TWHya,$\eta$Cha,UCC,UCL,Isolated\\
         Structured & 1.3$\pm$0.1 & 1.0$\pm$0.2 &0.1$\pm$0.02&all\\         
         \hline
    \end{tabular}
\end{table}

Figure \ref{fig:mdiskmstar} presents the relation between disk dust mass and stellar mass for our sample, which is grouped into the three age categories young, intermediate and old. Transition and ring disks (structured disks) are marked. The correlation for each subsample (excluding the structured disks) is computed using the linear regression procedure of \citet{Kelly2007} with the \textit{linmix} python package. This procedure takes into account upper limits and intrinsic scatter in the data. The best-fit parameters are given in Table \ref{tbl:fit}, with intercept $\alpha$, slope $\beta$ and variance on the scatter $\delta$, following \citet{Ansdell2017}. The slope for the intermediate age group is flatter than expected, which is likely due to the wider range of ages \citep{Winter2019} compared to the other groups. 
A similar trend is reproduced as in previous works \citep{Ansdell2017,Pinilla2018,Pinilla2020}: the dust mass and stellar mass are correlated, and the dust mass drops more significantly with age for the lower mass stars. However, the structured disks show a flatter relation, although with significant scatter of the extended disks. Furthermore, structured disks appear to be more common around higher mass stars based on visual inspection of the number of data points in this plot, which is further explored in the next section.

\begin{figure}
    \centering
    \includegraphics[width=\textwidth]{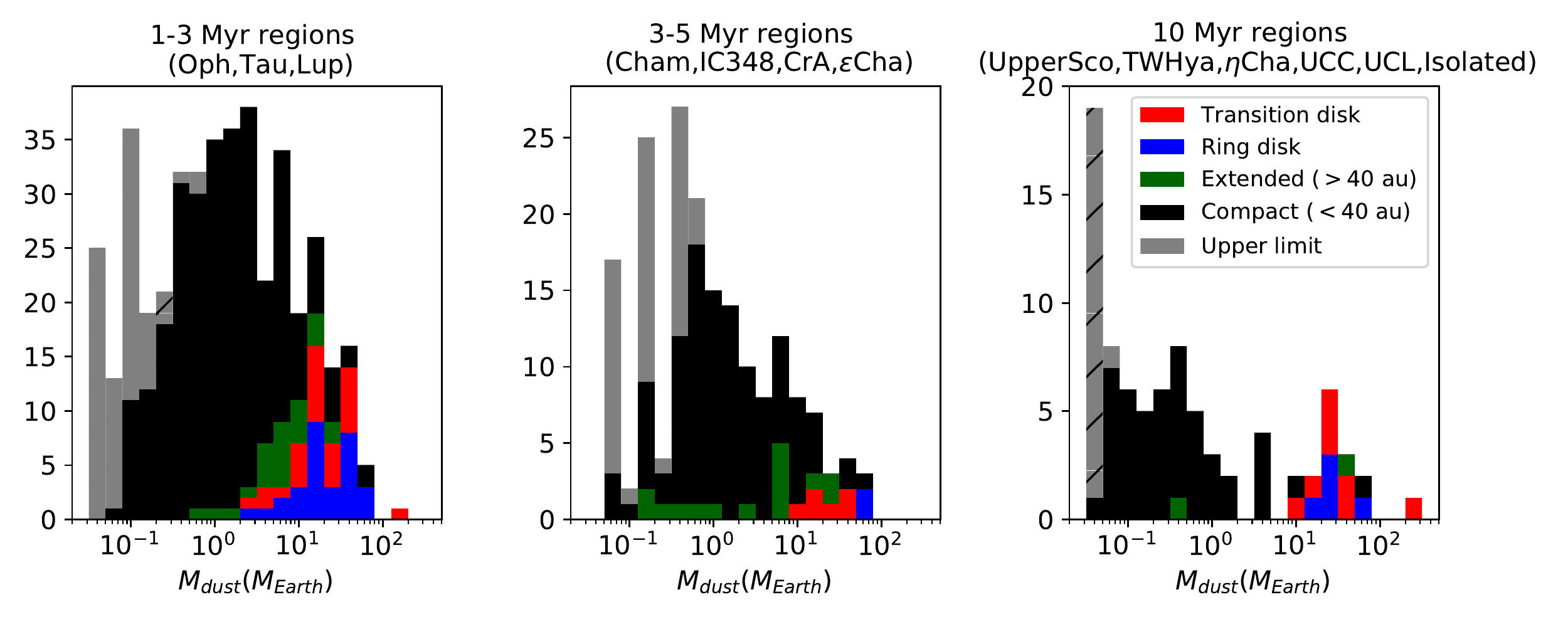}
    \caption{Disk dust mass distribution, split up by age group. Structured disks are marked in red (transition disks), blue (ring disks) and green (extended disks), respectively, whereas upper limits are marked in grey and disks with unknown structure in black. 
    The distribution shows a clear drop in dust masses of nonstructured disks with age, whereas structured disks remain in the higher ($\gtrsim10 M_{\rm Earth}$) regime.}
    \label{fig:massdistribution}
\end{figure}

In Figure \ref{fig:massdistribution} the dust mass distribution in each age group is shown again, but now as histogram. Transition, ring and extended disks are marked in red, blue and green, respectively, whereas compact disks are black and dust mass upper limits are indicated in grey. 

\begin{figure}
    \centering
    \includegraphics[width=\textwidth]{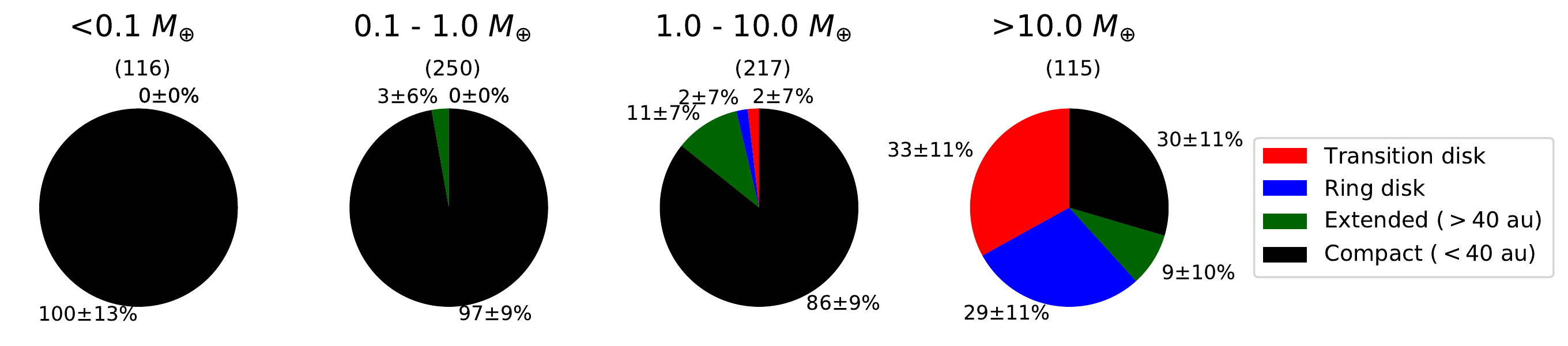}
    \caption{Distribution of disk morphologies for different disk dust mass bins. Different disk morphologies are marked in red (transition disk), blue (ring disk), green (extended disk) and black (compact). The numbers in brackets above each pie chart indicate the number of targets in that dust mass bin.}
    \label{fig:dustmasspie}
\end{figure}

The distribution plot show that structured disks remain located in the $\gtrsim 10 M_{\rm Earth}$ dust mass regime at any age, whereas the bulk of disks (without large scale substructure) has a lower dust mass distribution ($\lesssim 1 M_{\rm Earth}$) which decreases with age. Although dust opacity may change as function of age as well, previous works have demonstrated that the dust disk size decreases with age for the majority of disks \citep[e.g.][]{Hendler2020}. Also, a change in dust opacity cannot explain the two separate evolutionary pathways observed in this figure.

Figure \ref{fig:dustmasspie} directly shows the fractions of transition and ring disks in different dust mass bins. 52 upper limits are above the 0.1 $M_{\rm Earth}$ threshold and included in the 0.1-1 $M_{\rm Earth}$ bin, although they could actually belong to the 0.1 $M_{\rm Earth}$ bin. Considering the large number of targets, this does not make a difference for the computed fractions within the error bars. This figure confirms the increased number of structured disks in higher disk mass regimes, as expected from predictions from dust evolution models, which show that large dust masses can only exist in the presence of pressure bumps which limit radial drift \citep[e.g.][]{Birnstiel2010,Pinilla2012a,Pinilla2020}. In particular,  at least two thirds of the disks in the high disk mass regime ($\gtrsim 10 M_{\rm Earth}$) are structured.

\begin{figure}[!ht]
    \centering
    \includegraphics[width=\textwidth]{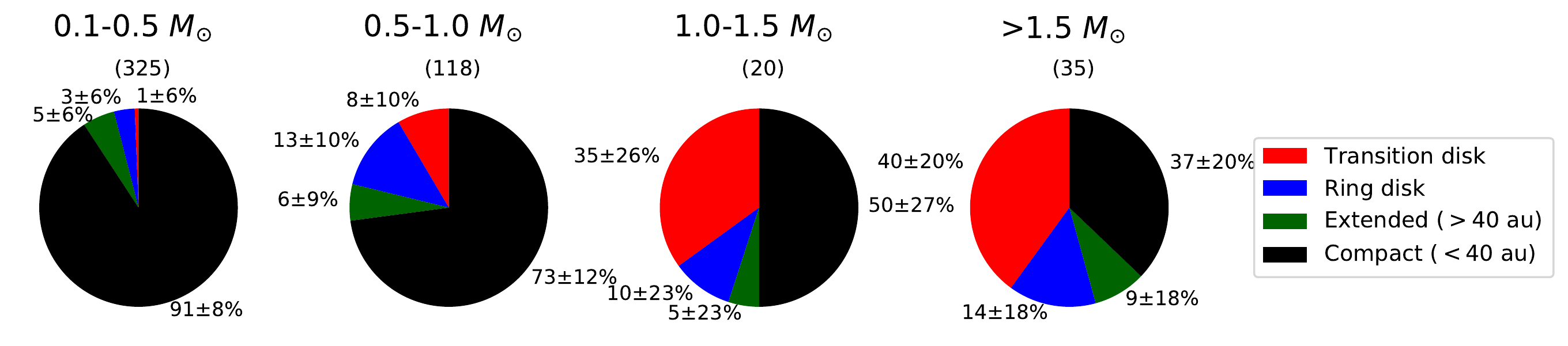}
    \caption{Distribution of disk morphologies for different stellar mass bins. Different disk morphologies are marked in red (transition disk), blue (ring disk), green (extended disks) and black (compact). The numbers in brackets above each pie chart indicate the number of targets in that stellar mass bin.}
    \label{fig:mstarmorphology}
\end{figure}

\begin{figure}[!ht]
    \centering
    \includegraphics[width=0.45\textwidth]{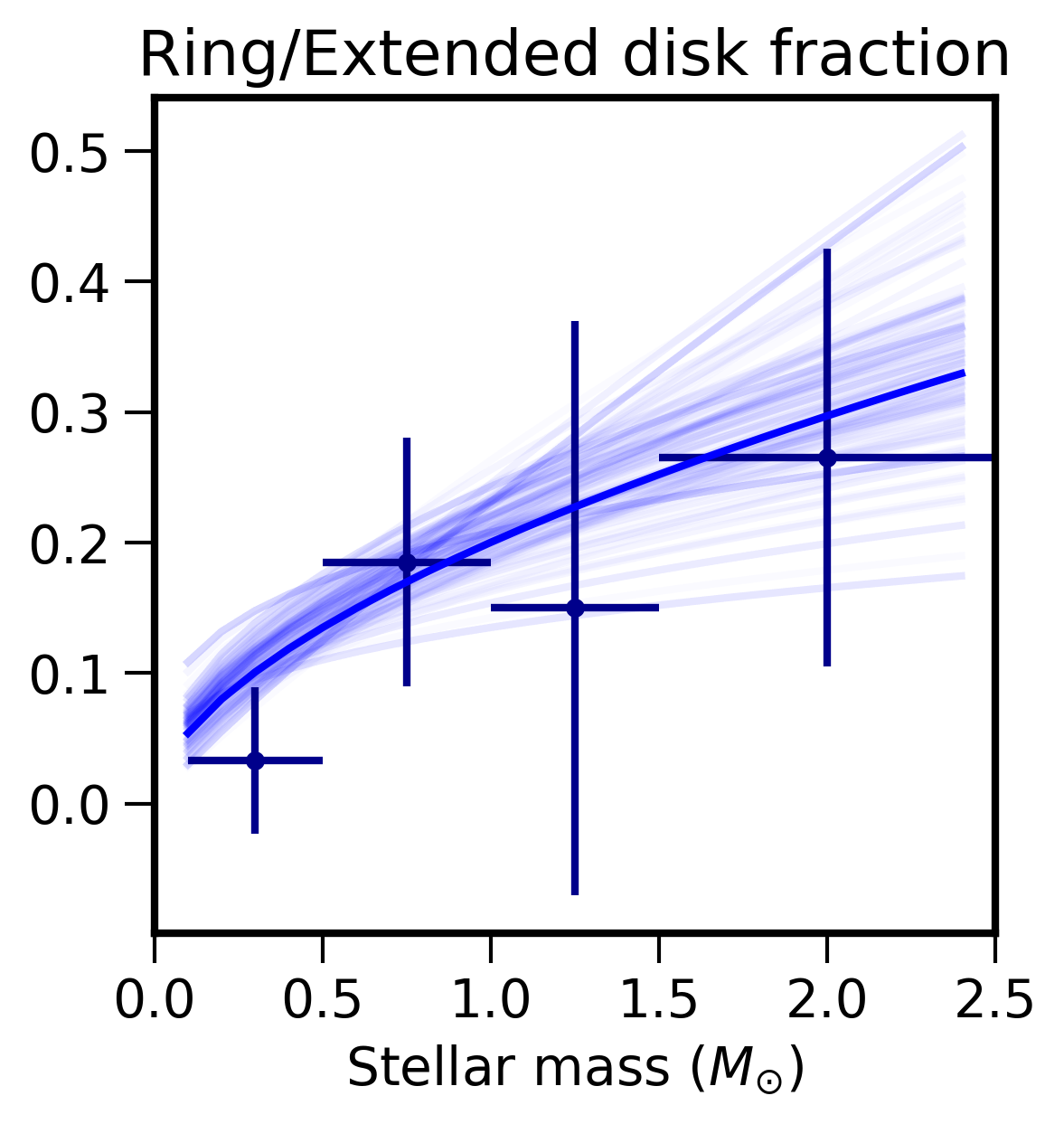}
    \includegraphics[width=0.45\textwidth]{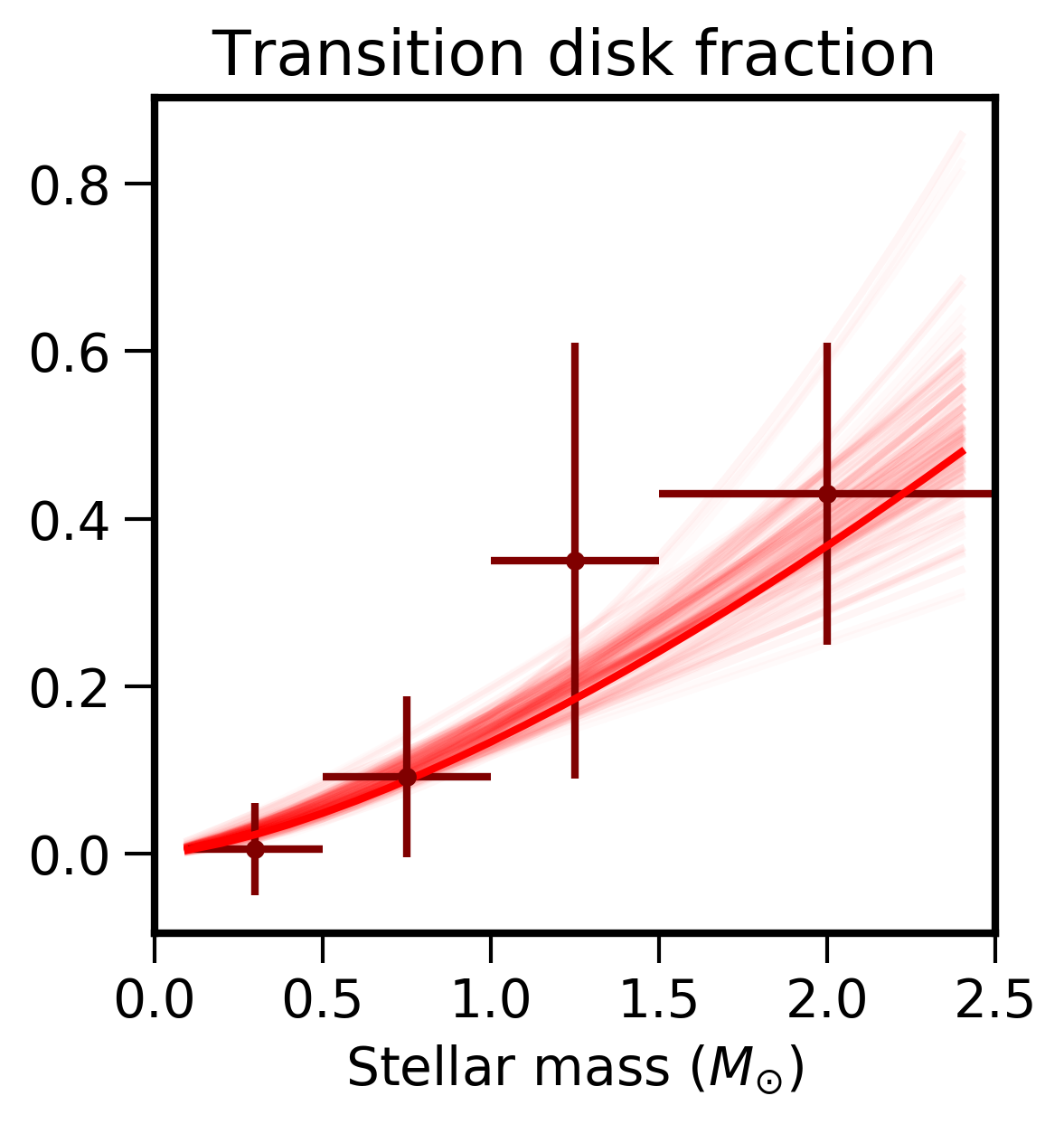}
    \caption{Comparison between the computed disk fractions of the disk morphology classifications transition disk and ring/extended disk in stellar mass bins (data points with error bars) and the disk fraction estimated from a power-law fit (solid lines) following Eqn. \ref{eqn:pl}. The transparent lines indicate the spread in MCMC samples from the fit.}
    \label{fig:testbins}
\end{figure}

Second, we explore the distribution of different types of disk structure with respect to stellar mass, regardless of the disk dust mass or age. Figure \ref{fig:mstarmorphology} presents pie charts of the distribution of disk morphologies across each stellar mass bin.

In order to check whether these fractions depend on the choice of mass bin, we compute the disk fractions directly from the disk populations assuming a power-law, following the Bayesian approach by \citet{Johnson2010}:
\begin{equation}
\label{eqn:pl}
    f(M_*) = C\cdot{M_*^{\alpha}}
\end{equation}
Using MCMC, we find the following best-fit values for transition disks ($C=0.13\pm0.03$,$\alpha=1.46\pm0.19$) and ring/extended disks ($C=0.20\pm0.02$,$\alpha=0.57\pm0.14$), respectively. Ring and extended disks are considered to be in the same category here, under the assumption that extended disks must contain pressure bumps to be large \citep{Pinilla2020,Long2020}. Figure \ref{fig:testbins} shows the fractions in bins and the best-fit power-law. Uncertainties on the binned values in Figure \ref{fig:mstarmorphology} are Poissonian and thus dependent on the number of objects in each stellar mass bin. The fits demonstrate that the binned fractions are reasonable estimates. 

For comparison, the best-fit values for a distribution without stellar mass dependence ($\alpha$=0) results in $C=0.13\pm0.16$ and $C=0.21\pm0.24$ for transition and ring/extended disks respectively. The Bayesian Information Criterion (BIC) test is used to determine which fit is statistically better. BIC is a criterion for model selection among a set of models in which the model with the lowest BIC is preferred. The BIC is formally defined as BIC = $k\ln(N)-2\ln(\hat{L})$, with $k$ the number of variables, $N$ the number of observations and $\hat{L}$ the maximized value of the likelihood function of the model. The BIC scores are found to be 208 and 384 for transition and ring/extended disk fraction for the stellar mass dependence model, and 272 and 396 for the model without stellar mass dependence, respectively. The stellar mass dependence is thus a statistically significant better fit, despite the increased complexity of the model. 

Figure \ref{fig:mstarmorphology} and \ref{fig:testbins} reveal that structured disks are much more common around intermediate mass stars than low-mass stars. 
The trend with transition disks is particularly strong: more than 40\% of the intermediate mass stars has a transition disk, compared to only 9\% of the sub-Solar stars and less than 1\% for the very low-mass stars. On the other hand, this also shows that transition disks are not unique for higher mass stars, just much less likely to be found around low-mass stars, albeit with large error bars. Within the assumption that the extended disks are essentially ring disks as well, the fraction of ring disks also increases with stellar mass, although not as significantly as the transition disks: from 8\% to 26\% from low to intermediate mass. Interestingly, the fraction of transition disks is higher than that of the ring disks for the super-Solar mass bins, but lower for the sub-Solar mass bins.

\section{Comparison with Exoplanet Statistics}
\label{sct:exoplanets}

In this section we compare the observed incidence of transitional and ringed disks around pre-main-sequence stars with the demographics of exoplanets around main-sequence and evolved stars. 
The stellar mass dependence of different types of exoplanets have been previously established, see e.g. \cite{Mulders2018} for a review. 
Of most relevance here, radial velocity surveys have established that Jupiter-mass planets are positively correlated with stellar mass \citep{Johnson2010,Reffert2015}, while the \emph{Kepler} transit survey discovered that super-Earths are anti-correlated with stellar mass \citep{Howard2012,Mulders2015a}.
We use these broader population trends to construct a demographical model of exoplanet occurrence rates that can be directly compared to the observed disk fractions.
In this comparison we assume that the gap locations are not representative for the exoplanet orbital radii. Inward migration can account for gaps at tens of au caused by planets that end up at 2-3 au.

We first explore the hypothesis that disk structures are created by giant planets in Section \ref{sct:rv}, and then in Section \ref{sct:kepler} we explore the hypothesis that transiting sub-Neptunes form in disks without structure.
Such a comparison between disks and exoplanets requires a careful consideration of detection biases in observed exoplanet samples (Mulders et al. subm.). 
We construct a demographical model of planets around stars of different masses based on measured planet occurrence rates from radial velocity and transit surveys, and use that as a basis to predict what fraction of observed protoplanetary disks would be transitional or have substructure. These planet occurrence rates are corrected for the varying sensitivity of exoplanets surveys to planets of different mass and around different types of stars, and reflect the intrinsic population of planets, though only within the detection limits of those surveys. 

\subsection{Radial Velocity Giant Planets}
\label{sct:rv}

\begin{figure}
    \centering
    \includegraphics[width=0.5\textwidth]{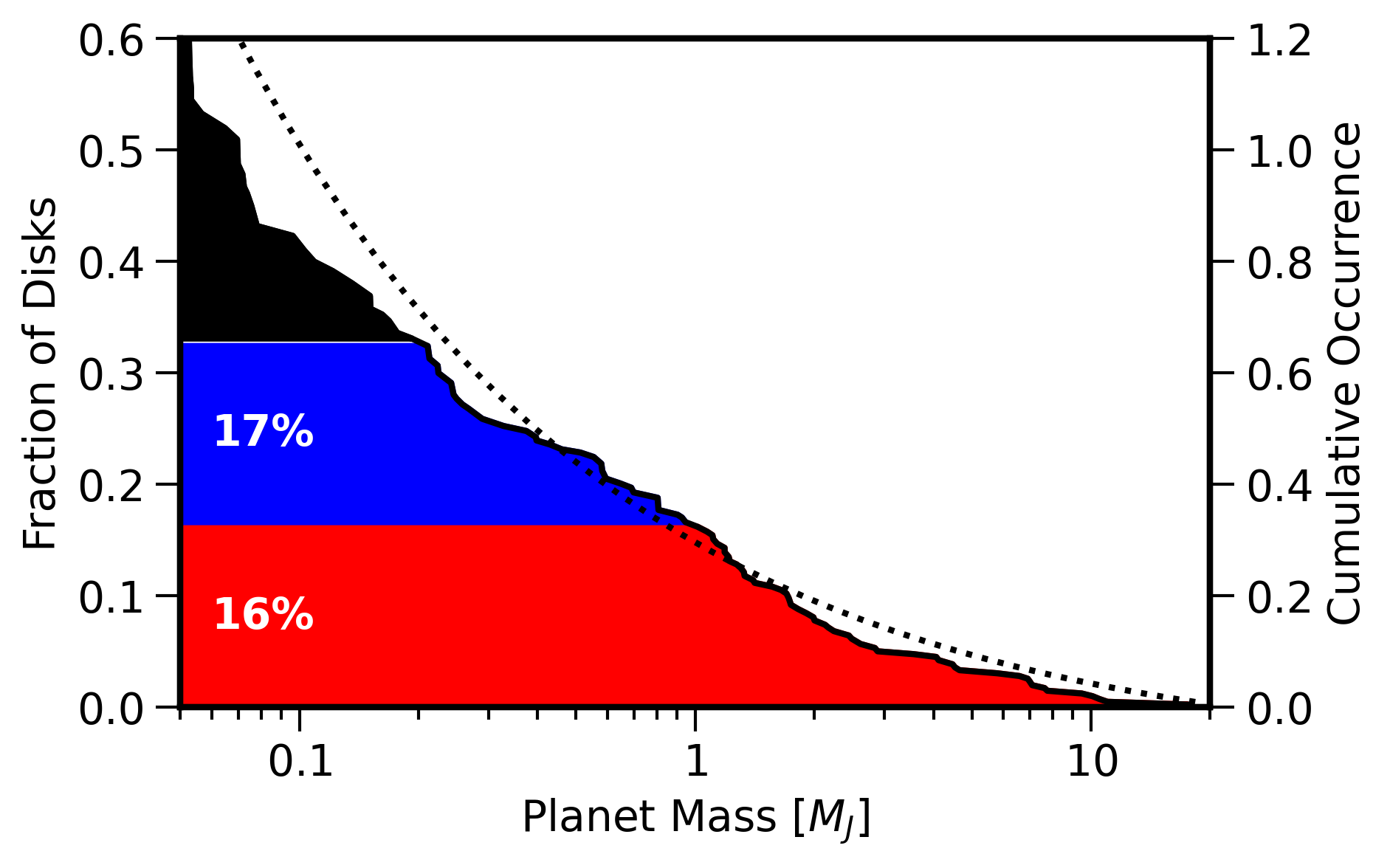}
    \caption{
    Giant planet occurrence rate based on the \cite{Mayor2011} radial velocity survey of sun-like stars, expressed as an inverse cumulative distribution. 
    The red region indicates what the fraction of transition disks would be if a Jupiter-mass planet can open a wide cavity in the disk. 
    The blue region indicates the what fraction of disks with structure would be assuming a $0.2 M_{\rm Jup}$ planet is massive enough to open a gap in the disk.
    The dotted line shows the integrated planet occurrence rate from the parametric model of \cite{Fernandes2019} and Eq. \ref{eq:exomodel}.
    }
    \label{fig:rv}
\end{figure}

No single exoplanet survey spans the same range of stellar mass as the protoplanetary disk sample described above, and also has enough sensitivity to detect planets over a wide range of planet masses and semi-major axes. Thus, we construct a demographical model from radial velocity-detected giant planets by combining the planet mass and semi-major axis distribution around solar-mass stars from \cite{Mayor2011} and \cite{Fernandes2019} with the stellar-mass dependence from \cite{Johnson2010}. The latter derived occurrence rates of Jupiter-mass planets around three samples of stars: low mass M dwarfs, solar mass stars, and higher mass stars that have evolved of the main sequence. While the stellar mass of this sample of "retired A stars" is somewhat uncertain, the identification of these stars as significantly more massive than solar ($M_\star >1.5 M_\odot$, and thus in our highest mass bin) appears secure \citep{Ghezzi2018,Malla2020}.

To estimate the fraction of stars with an exoplanet massive enough to open up a wide cavity (transition disk) or an annular gap (ringed or extended disk) we start with the occurrence rate distribution of giant planets around sun-like stars measured by \cite{Mayor2011}, because that survey extends into the Neptune-mass regime. Figure \ref{fig:rv} shows the inverse cumulative distribution, e.g. the integrated planet occurrence rate of planets more massive than a certain value. 
The left axis shows how this translates to the fraction of disks with an exoplanet above that mass. This is essentially a re-normalization of the y axis that assumes there is one giant planet per planetary system and that exoplanets occur only around stars that also have disks, which is here assumed to be 50\%. The latter assumption compensates for the typical infrared disk lifetime of a few Myr \citep{Mamajek2009} and the inclusion of young and old star forming regions in our disk sample. We note that the infrared disk lifetime (`presence of small grains') is not equivalent to the presence of a massive millimeter dust disk (`presence of large grains') as these dust populations evolve separately \citep{Sellek2020}.

Only sufficiently massive giant planets are able to open up a wide enough cavity to create a transition disk. We can therefore estimate the expected fraction of transition disks from the number of stars that have an exoplanet more massive than a certain mass threshold, $M_\text{trans}$.
The red area shows that the estimated fraction of cavity-opening planets based on a threshold mass of $M_\text{trans}= 1 M_{\rm Jup}$, which is 16.3\% for the sample of sun-like stars. 

Planets that are not massive enough to create a transition disk but massive enough to create gaps with pressure bumps at its edges in the disk would create ringed dust disks in high-resolution observations.
We estimate the fraction of gapped disks from the number of stars that have an exoplanet more massive than $M_\text{ring}$, but less massive than $M_\text{trans}$. 
The blue area shows the estimated fraction of disks with a planet more massive than $M_\text{ring}= 0.2 M_{\rm Jup}$ is 16.5\% and the black area the fraction of disks with a planet $<0.2 M_{\rm Jup}$ or no planet.

Now we extend the demographic model to stars of lower and higher masses to compare with Figure \ref{fig:mstarmorphology}.
We add a stellar mass dependence to the parametric planet population model of \cite{Fernandes2019}:
\begin{equation}\label{eq:exomodel}
\frac{d^2 f(M_p,M_\star) }{d M_p d M_\star} = C M_p^a M_\star^b,
\end{equation}
where $a= -0.45$ from \cite{Fernandes2019}, $b=1$ from \cite{Johnson2010}, and $C$ a normalization constant such that $\int_{M_p>M_J} f(M_p,M_\odot) d M_p = 6.2\%$ again following \cite{Fernandes2019}.
We note that this functional form implicitly assumes that the stellar mass dependence measured for Jovian mass planets also applies to lower mass planets, though this has not been observationally constrained. We will test this assumption below.

We omit the semi-major axis dependence because the bulk of the giant planet are located near the peak at 2-3 au, and the population model contains only a small number of planets outside of the detection limits of ~10 au. 
The dotted line in Figure  \ref{fig:rv} shows that this parametric occurrence rate model provides a good match to the occurrence rates from \cite{Mayor2011} around solar-mass stars.
We therefore do not expect that exoplanets at large separations but below the detection limits of direct imaging surveys would provide a significant contribution to the total number of giant planets.  

\begin{figure}
    \centering
    \includegraphics[width=\textwidth]{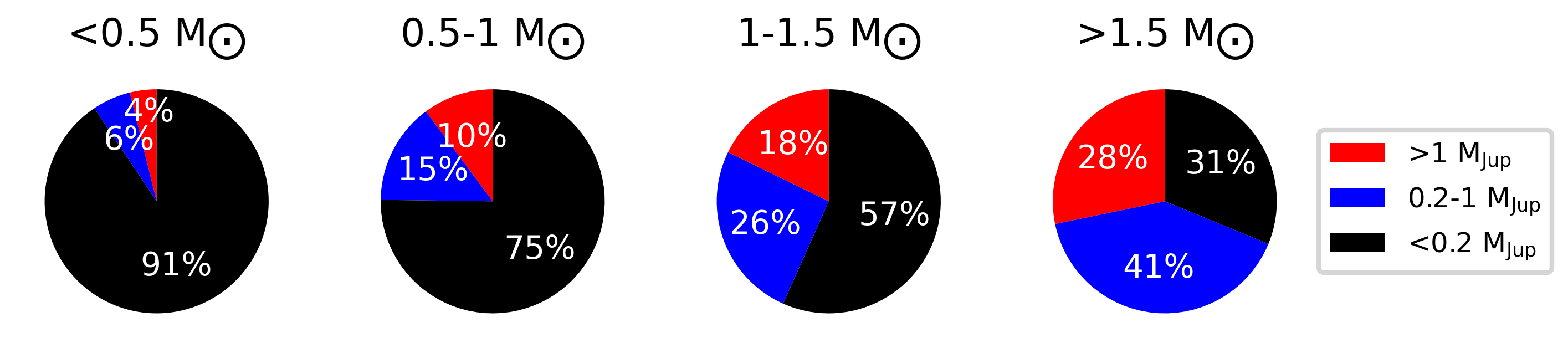}
    \caption{
     Predicted fractions of protoplanetary disks hosting an exoplanet massive enough to create a transition disk cavity ($>1 M_{\rm Jup}$, red) or only massive enough to create a narrow annular gap ($>0.2 M_{\rm Jup}$, blue). The fractions are calculated for the different stellar mass bins using the exoplanet mass distribution from \cite{Fernandes2019} and the stellar-mass dependence from \cite{Johnson2010}, see text and Equation \ref{eq:exomodel}. 
The predicted fraction of structured disks is consistent with the observed disks, but the transition disks are over(under) predicted in the lowest (highest) stellar mass bin.
    }
    \label{fig:mstarplanets}
\end{figure}

Finally, we calculate the probability that each star in the disk sample has a cavity or a gap based on its stellar mass and the occurrence of exoplanets above the mass thresholds $M_\text{trans}$ and $M_\text{ring}$. 
Figure \ref{fig:mstarplanets} shows the predicted number of transition and gapped disks for the same stellar mass bins as figure \ref{fig:mstarmorphology}, assuming a cavity-opening mass of 1 Jupiter mass and a gap-opening mass of 0.2 Jupiter mass.
Figure \ref{fig:mstarplanets} shows a similar trend as the morphologies in Figure \ref{fig:mstarmorphology}: Super-Jovian planets ($>1 M_{\rm Jup}$, red) are much more common around higher mass stars than lower-mass stars, but the predicted trend for the number of transition disks is weaker than the one observed. 
The predicted fraction of structured disks (with either a gap or cavity) matches the observed trend using $M_\text{ring} = 0.2 M_{\rm Jup}$ (blue).

\begin{figure}
    \centering
    \includegraphics[width=0.49\textwidth]{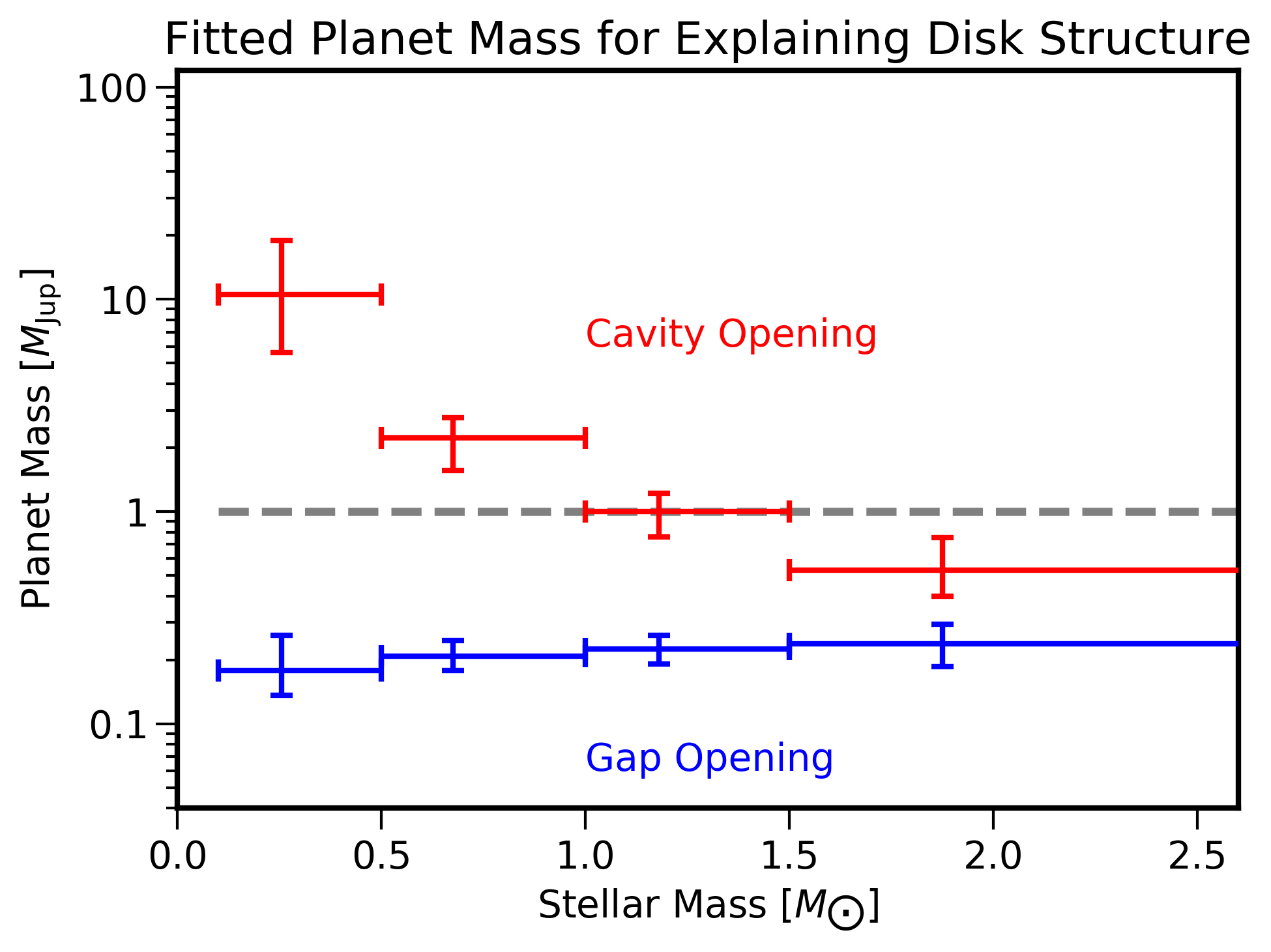}
    \includegraphics[width=0.49\textwidth]{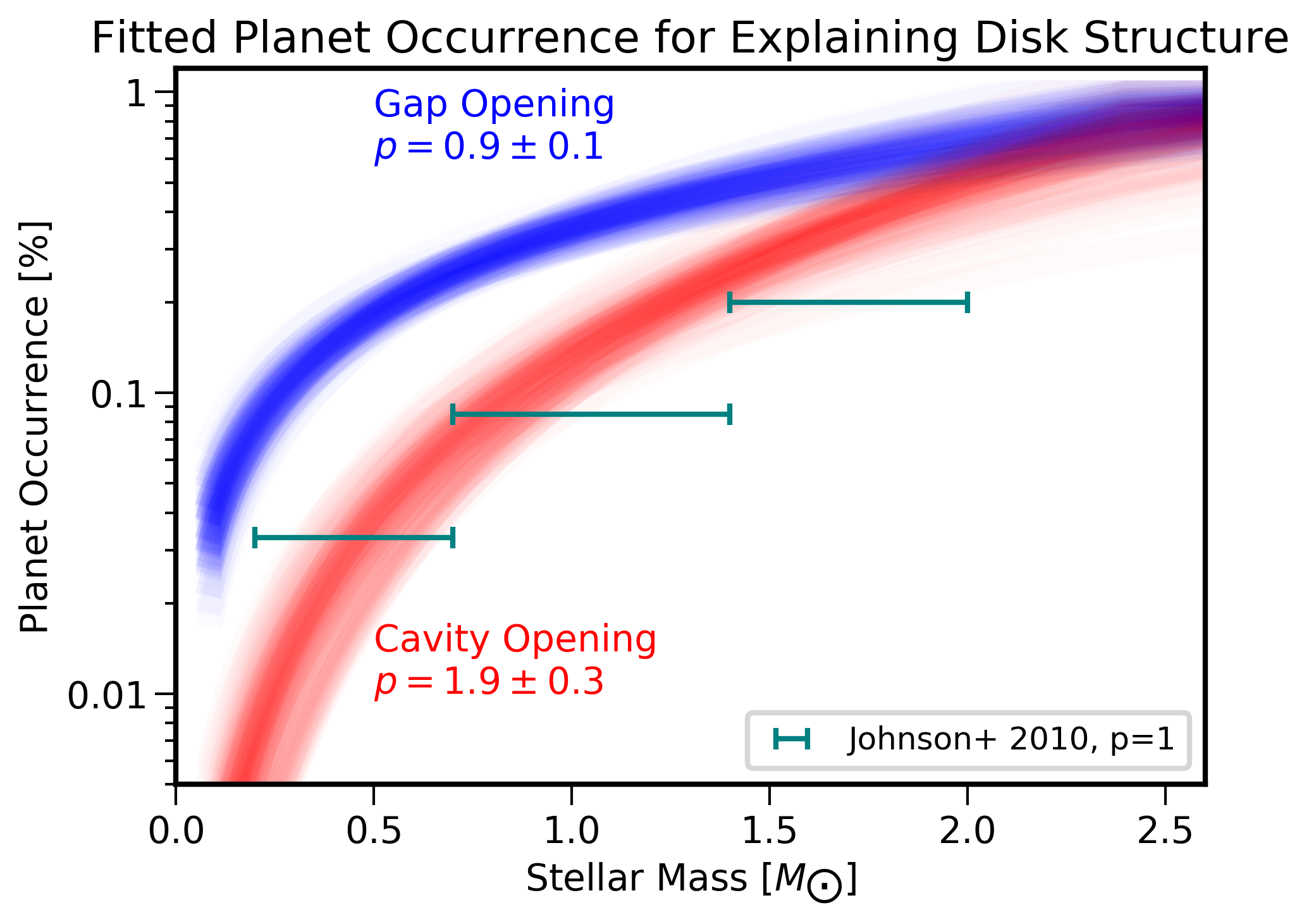}
    \caption{
    Two possible modifications to the exoplanet population model that can explain the low fraction of transition disks around M dwarfs. 
    Left: An increased minimum exoplanet mass required to create a cavity in the disk. 
    The exoplanet mass threshold for creating structured disks is consistent with being independent of stellar mass. The threshold exoplanet mass for creating a transition disks is consistent with a linear anti-correlation ($M_p \propto M_\star^{-1.3\pm0.3}$). 
    Right: a steeper stellar-mass dependence for cavity-opening planets than for gap-opening planets.
    }
    \label{fig:scenarios}
\end{figure}

\begin{figure}
    \centering
    \includegraphics[width=\textwidth]{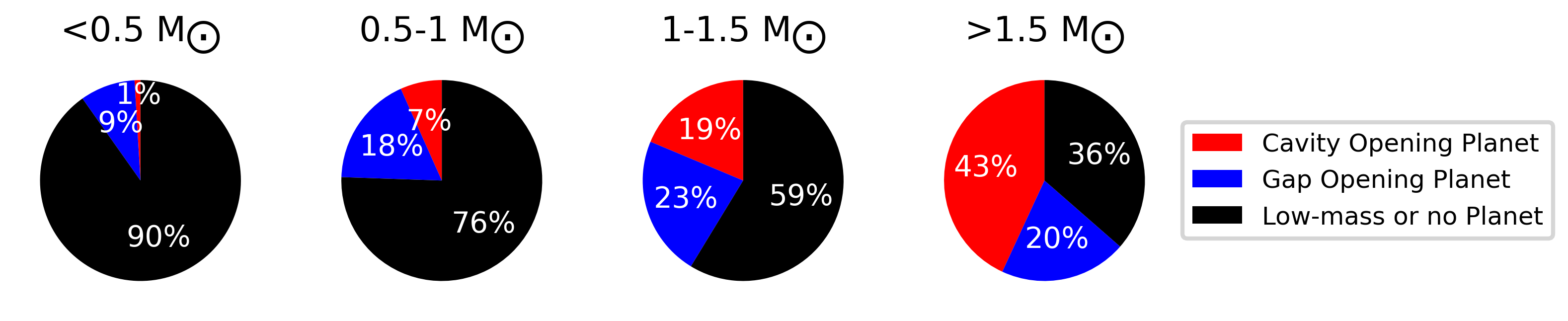}
    \caption{
    Predicted fractions of protoplanetary disks hosting a cavity opening or gap opening planet (as Figure \ref{fig:mstarplanets}), using the best-fit MLE solution from figure \ref{fig:scenarios} 
    }
    \label{fig:mstarplanetratios}
\end{figure}

To get a better match to the observed fractions of transition disks, there are two possibilities: varying the threshold mass for a planet to open a cavity, or adjusting the assumed scaling of planet occurrence rate with stellar mass from equation \ref{eq:exomodel}. 

First, we allow the threshold planet masses $M_\text{trans}$ and $M_\text{ring}$ to vary with stellar mass. We parameterize this dependence as a power-law in stellar mass, $M_\text{trans}= M_p (M_\star/M_\odot)^q$. We then do a maximum likelihood estimate using \texttt{emcee} \citep{DFM2013}, where we vary $M_p$ and $q$ to match the observed fraction of transition disks at each stellar mass. The likelihood is calculated using binomial statistics as in \cite{Johnson2010}.
Figure \ref{fig:scenarios}, left panel, shows the estimated mass thresholds for a planet opening a gap (blue) or cavity (red). The binned fractions of the best-fit model are also presented in Figure \ref{fig:mstarplanetratios}. The MLE fit has a steep anticorrelation between the threshold planet mass and stellar mass, 
\[
M_{p,\text{trans}} = 1.16_{-0.24}^{+0.30},
q_\text{trans} = -1.36_{-0.33}^{+0.35}
\]
which provides a better match to the transition disk fraction in the lowest and highest stellar mass bin. 
The threshold mass for a ringed disk is consistent with having no stellar mass dependence:
\[
M_{p,\text{ring}} = 0.21_{-0.03}^{+0.04}, 
q_\text{ring} = 0.12_{-0.21}^{+0.21}
\]

It is not clear why M dwarfs would require more massive planets to open a cavity. 
If cavities are indeed caused by eccentric companions as suggested by \citet{Muley2019, vanderMarel2020}, the planet/star ratio required for eccentricity is constant and thus lower mass stars would require lower mass planets.
Therefore, we also explore a second scenario next where we keep the threshold mass fixed but fit for the stellar mass dependence of the planet occurrence rate, $p$. 

The right panel of Figure \ref{fig:scenarios} shows the estimated stellar mass dependence of the giant planet occurrence rate that would match the fractions of structured disks. The occurrence rate of gap-opening planets ($b=0.9\pm0.1$) is consistent with the linear stellar mass dependence from \cite{Johnson2010} as previously assumed. The occurrence rate of cavity opening planets would have to depend more strongly on stellar mass ($b=1.9\pm0.3$) than is observed for giant exoplanets. 
The binned fractions are statistically indistinguishable to those of the other scenario presented in Figure \ref{fig:mstarplanetratios}. To discriminate between the two scenarios, a larger exoplanet sample is needed, in particular one probing the planet mass distribution around M dwarfs, including sub-Jovian planets.

\subsection{Transiting Close-in Planets}
\label{sct:kepler}
The occurrence rates of transiting planets smaller than Neptune, commonly referred to as super-Earths, are known to be anti-correlated with stellar mass \citep{Howard2012,Mulders2015a} and increase from spectral types F through M in the \emph{Kepler} survey. Recently, this increase in planet occurrence rate has been shown to correspond to an increase in the fraction of stars with planetary systems as well \citep{Yang2020,He2020}. 
Thus, it is not likely that these planets cause or form directly in the observed disk gaps and cavities, which are positively correlated with stellar mass. Instead, we explore the opposite hypothesis that these structures, and the associated giant planets, inhibit the formation or inward migration of the super-Earths that are detectable with \emph{Kepler}. 
This would give rise to an anti-correlation between the fraction of disks with gaps and the fraction of stars with transiting exoplanets.

There is some theoretical motivation for this hypothesis, motivated by the lack of super-Earths in the Solar System and linking this to the presence of Jupiter. For example, \cite{Izidoro2015} hypothesized that Jupiter may have prevented the inward migration of super-Earths. Along similar lines, \cite{Lambrechts2019} found that super-Earth formation is dependent on the pebble flux, and that Jupiter may have starved the inner solar system from pebbles. Giant planets such as Neptune and Uranus are expected to limit pebble drift as well. Such mechanisms, if they also operate in the disks with observed cavities, would reduce the occurrence rates of close-in super-Earths. In this section, we test if the anti-correlation between close-in super-Earths and disk structures predicted by such mechanisms could be consistent with the Kepler planet occurrences rates.

\begin{figure}
    \centering
    \includegraphics[width=0.8\textwidth]{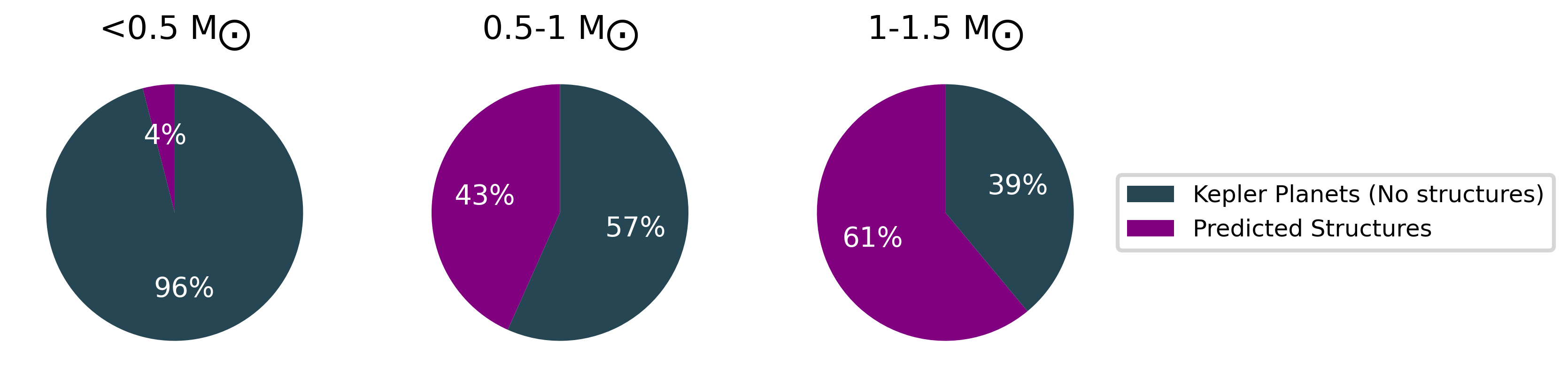}
    \caption{
    Predicted fraction of structured disks based on \emph{Kepler} exoplanet statistics. 
    We assume here that a disk with a cavity or gap (purple) would not form planets detectable with Kepler (green), and vice versa. 
    The \emph{Kepler} planet occurrence rates are only measured for M,K,G, and F stars, and thus a prediction for the highest stellar mass bin is missing. The stellar mass dependence from \emph{Kepler} is stronger than what is observed in disks, and under(over) predicts the fraction of structured disks in the lower (higher) stellar mass bin.
    }
    \label{fig:kepler}
\end{figure}

We calculate the fraction of stars with close-in planetary systems, $F$, from the \emph{Kepler} planet occurrence rates, $\eta$ and an estimate for the number of planets per planetary system, $\bar{n}$.
\[
F = \bar{n}\,\eta
\]
We calculate the planet occurrence rate, $\eta$, for different stellar mass bins as described in \citep{Mulders2018epos}, using the \emph{Kepler} \texttt{DR25} planet candidate catalogue \citep{Thompson2018} and survey completeness \citep{Burke2017}.
The highest stellar mass bin ($>1.5 M_\odot$) does not contain enough stars and planets to calculate an occurrence rate for sub-Neptunes. 
The fraction of stars with planets, $\bar{n}$, is in the range of 2-6 based on forward modeling of planet populations \citep[e.g.][]{Zhu2018,Mulders2018epos,He2019}.
Here we will assume $\bar{n} = 4.5$, mainly such that the fraction of stars with planets, $F$, does not exceed 1. 
Figure \ref{fig:kepler} shows the fraction of stars with Kepler planetary systems and the predicted fraction of structured disks, equal to $1-F$.
The model under-predicts the number of structured disks for M dwarfs but over-predicts that same number for the more massive stars. 

\begin{figure}
    \centering
    \includegraphics[width=0.8\textwidth]{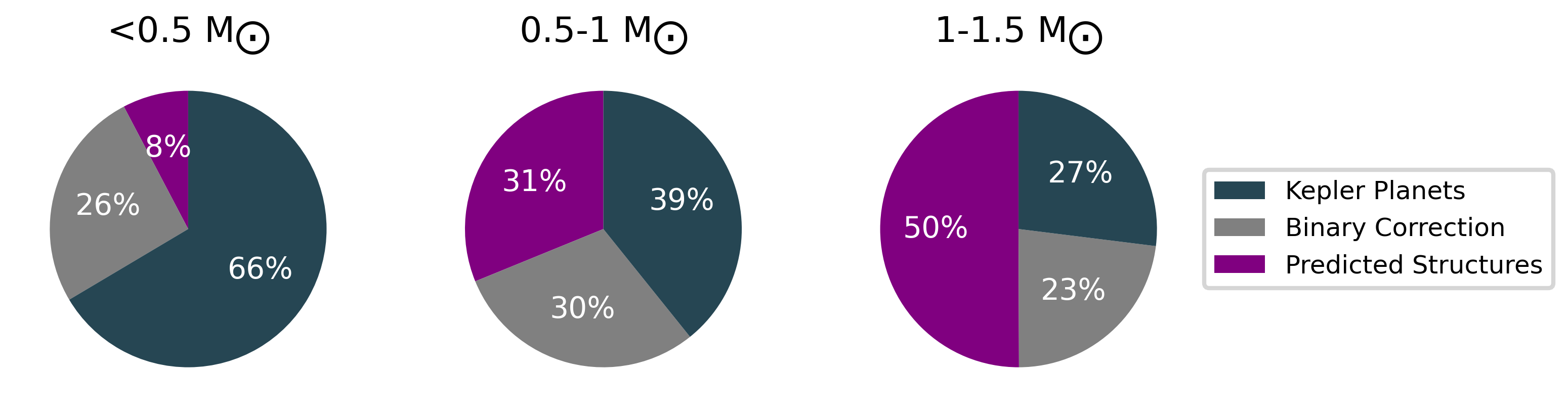}
    \caption{
    Same as Figure \ref{fig:kepler}, but accounting for the stellar binary fraction in the Kepler survey following \cite{MoeKratter2019}. 
    The estimated fraction of protoplanetary disks without Kepler planets is consistent with the fraction of disks with observed structure (cavity or gap).
    }
    \label{fig:kepler-binaries}
\end{figure}

Planet occurrence rates for single stars are underestimated by the presence of binaries in the \emph{Kepler} sample. \cite{MoeKratter2019} showed that accounting for binaries can explain up to half of the known stellar-mass dependence for sub-Neptunes. Because the disk sample is biased towards single stars due to the infrared selection criteria \citep[][and Section \ref{sct:binaries}]{Kraus2012} we apply this binary correction in Figure \ref{fig:kepler-binaries}, where we also adjust the number of planets per system, $\bar{n}$, to $6.5$ to make sure that the fraction of M dwarfs with planets does not increase above 100\%. 
The number of stars with close-in planetary systems now roughly matches the observed incidence of disks without gaps or cavities in each of the stellar mass bins.
This comparison shows that the elevated occurrence rates of transiting planets around M dwarfs may be connected to those stars having fewer disk structures to impede the drift of pebbles or the migration of super-Earths.

\newpage
\section{Discussion}
\label{sct:discussion}

\begin{figure}[!ht]
    \centering
    \includegraphics[width=0.8\textwidth]{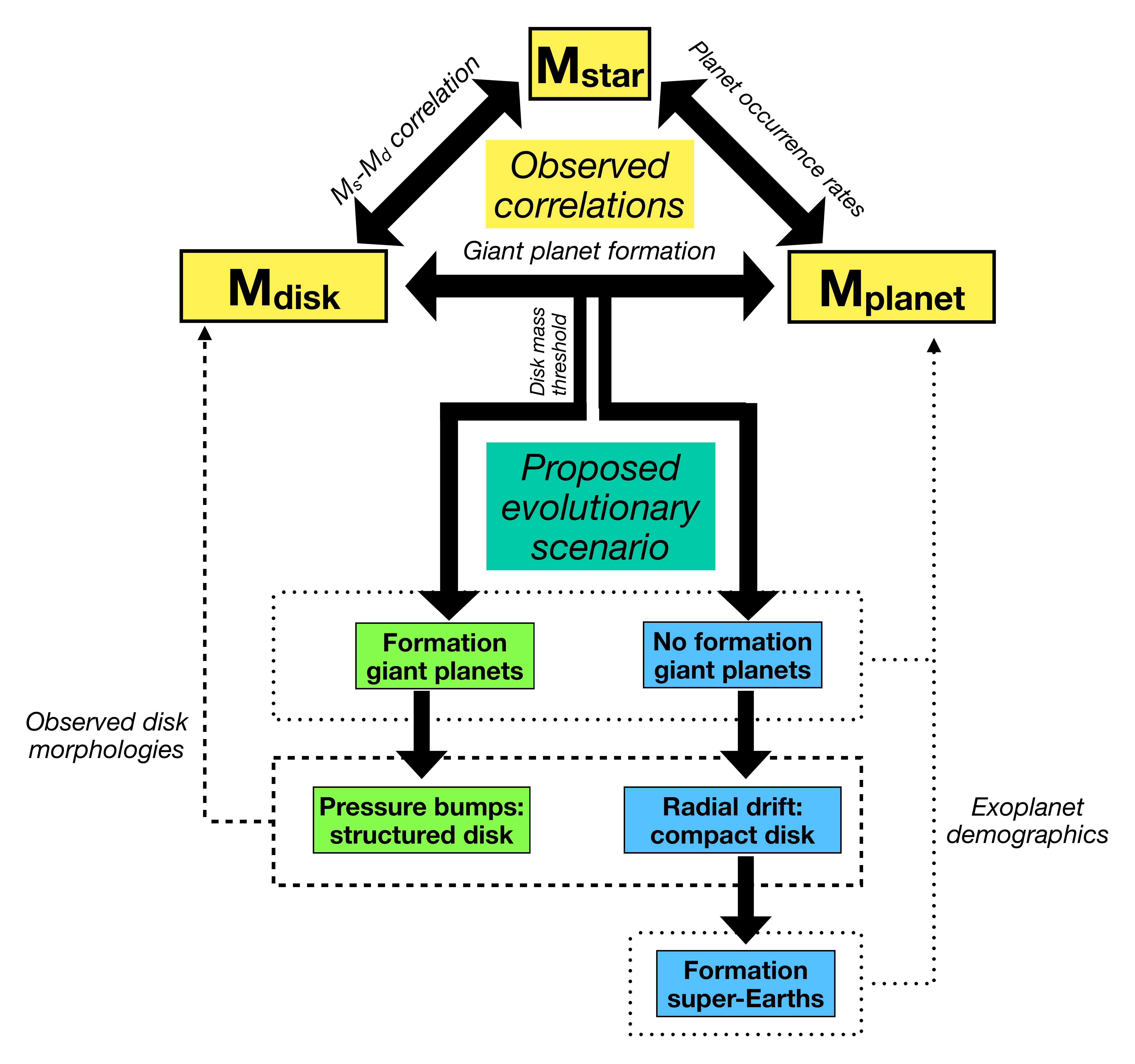}
    \caption{Flow chart describing our proposed scenario for disk evolution connected with disk and exoplanet demographics.}
    \label{fig:flowchart}
\end{figure}

Our work reveals that the fraction of large scale structured disks has a strong stellar mass dependence, which is similar to the stellar mass dependence of the occurrence rates of giant exoplanets, and the opposite of the occurrence rates of Kepler planets, which were both previously known \citep{Johnson2010,Howard2012}. With the assumption that structured disks are the result of pressure bumps at the edges of gaps carved by giant planets while the disk dust radius is otherwise regulated by radial drift, we propose a scenario to link these trends directly (see Figure \ref{fig:flowchart}). In addition, this scenario can explain the scatter in the size-luminosity relation \citep{Rosotti2019}, the steepening of the slope of the dust mass-stellar mass relation \citep{Ansdell2017}, the flat slope of structured disks in the dust mass-stellar mass relation \citep{Pinilla2018} and the presence of massive outlier disks in older regions \citep{Ansdell2020}, through a combination of radial drift and dust trapping in pressure bumps in disks around different stellar masses \citep{Pinilla2020}. Our scenario can be summarized as follows.

Disks form around young protostars with a range of gas masses. The initial gas disk mass determines whether giant planets can form in the outer disk. The mass of these planets then determine whether pressure bumps are created in the disk, and thus whether the disk remains massive in millimeter grains and become observable as transition/ring/extended disk with ALMA. If the planets are not massive enough, the millimeter grains drift inwards leading to low-mass, compact disks that could form Kepler-like planetary systems (super-Earths) instead. 
The capability of the disk to form giant planets thus determines its evolutionary path and millimeter-dust lifetime. Since higher mass stars are surrounded by higher mass dust disks \citep{Ansdell2017}, higher mass stars are also more likely to form giant planets, assuming a universal gas-to-dust ratio across stellar masses. This is an additional argument to support the stellar mass dependence of giant planet occurrence rates. A possible counterargument against this scenario would be that substructure remains hidden in the low-resolution observations of compact disks, which would remove the derived stellar mass dependence of structured disks. This is further discussed in Section \ref{sct:uncertainties}.

\subsection{Link between planets and disk structures}
\label{sct:planetstruct}
The similarities between the disk demographic fractions and exoplanet mass bin fractions as function of stellar mass (Figure \ref{fig:mstarmorphology} and \ref{fig:mstarplanets}) suggest a connection between the observed gaps in disks and recently formed planets, where the most massive planets ($\sim$Jupiter mass) are responsible for the carving of the widest gaps in transition disks, moderately massive planets ($\sim$Neptune - Jupiter mass) are responsible for the structures seen in ring disks, whereas lower mass planets below about a Neptune mass do not affect the disk structure and the disk dust structure is set by radial drift. The fact that both structured disk fraction and massive planet fraction show a strong positive correlation with stellar mass provides the most convincing evidence for a connection here. As massive stars are generally surrounded by more massive disks (Figure \ref{fig:mdiskmstar}), it is not surprising that they are also forming more massive planets. 

The majority of giant exoplanets are found at smaller orbital radii than the ALMA spatial resolution limit, and thus could not be responsible for creating the gaps in protoplanetary disks if they formed at their current locations. However, it is very likely that these planets have migrated inwards (Type I/II migration) during the lifetime of the disk due to the torques between disk and planet \citep{Paardekooper2010,Lodato2019} and we consider the current location of the exoplanets as less relevant for our conclusions. The potential low efficiency of planet formation at the large gap radii is further discussed in Section \ref{sct:planetformation}. 

The observed connection between disk morphology and giant planet fraction is consistent with pressure bumps at the edge of planet gaps, where the dust gets trapped and concentrated, resulting in rings, gaps and cavities \citep{Pinilla2012b}. Planet gap edges require a minimum planet mass to have sufficiently strong pressure bumps to trap the millimeter dust at large radii. This threshold was initially set at 20 $M_{\rm Earth}$ ($\sim$1.2 $M_{\rm Nep}$) for a small set of planet-disk interaction models at $\alpha=10^{-3}$ for a 1 $M_{\odot}$ star \citep{Rosotti2016}, but later works show that the threshold depends on e.g. stellar mass, scale height (location in the disk) and viscosity \citep[]{Sinclair2020} and the minimum planet mass encompasses a range of values in the super-Earth regime with a mass $\sim$10-20$M_{\rm Earth}$, or $\sim$0.6-1.2 $M_{\rm Nep}$. Interestingly, \citet{Sinclair2020} shows that sub-Solar stars require somewhat higher mass planets to create a pressure bump compared to Sun-like stars. We do not see evidence of such a reverse stellar-mass dependence in the estimated fraction of ring disks though.

Our results suggest that with a higher threshold of $\sim$0.2 $M_{\rm Jup}$ or $\sim$60 $M_{\rm Earth}$, there are already enough exoplanets to account for the observed number of disks with gaps. This allows for the possibility that not all planets create observable disk structures, for example because some planets are located too close to their host stars for their gaps to be observable. 
On the other hand, many ring disks contains multiple gaps  which are likely caused by multiple planets, whereas our exoplanet demographic model assumes a single giant planet per star. If multiple planets per disk are required to create the observed structures,that implies that the minimum planet mass must be lower. Furthermore, gaps in the inner 10 au of the disk would remain undetectable due to high optical depth and lower scale height \citep{Andrews2020}. An exact determination of the planet mass range responsible for ring disks is beyond the scope of this study, but even the current simple approach strongly suggests that there must be a minimum value in the (Super-)Neptune mass range for clearing gaps, which is consistent with the exploration studies in planet-disk interaction simulations \citep{Rosotti2016,Sinclair2020}. 

Second, large transition disk cavities are likely caused by the most massive giant planets, around a Jupiter mass and above, where the required minimum planet mass decreases with stellar mass in order to be consistent with the high transition disk fractions in the highest stellar mass bins. \citet{vanderMarel2020} proposed a scenario where transition disks must harbor planets that are so massive that their orbits become eccentric, clearing a much wider gap than a single planet would be capable of \citep{dAngelo2006,Muley2019}. This threshold however depends directly on the stellar mass and the requirement for eccentricity is set at $q>$0.003 \citep{kleyDirksen2006} implying a minimum planet mass requirement that instead increases with stellar mass and is a factor of a few higher than what is derived in our planet mass threshold fit (Figure \ref{fig:mstarplanetratios}). 

It is possible that this threshold holds and only some of the transition disks are caused by individual planetary companions, whereas the remainder are caused by close stellar companions, i.e. circumbinary disks \citep[e.g.][]{Harris2012} or multiple Jovian planets \citep{Dodson-Robinson2011,Close2020}. A binary companion can clear a cavity of 2-5 times the binary separation \citep{Artymowicz1994,Hirsh2020}, so for transition disk cavities of $\sim$25-100 au we are interested in binary companions at $\sim$10-50 au separation. This is further discussed in Section \ref{sct:binaries}. When multiple Jovian planets are responsible for the wide transition disk cavities, this would shift our planet mass threshold by a factor of a few, but this does not change our major conclusions.

\subsection{Disk dust evolution}
\label{sct:diskevolution}
The two observed evolutionary pathways in Figure \ref{fig:massdistribution} are consistent with our main picture of the disk dust evolution process dominating the disk dust size and morphology: whereas structured disks maintain their dust mass due to the presence of pressure bumps (and thus planets), the bulk of the disks (compact) decrease their dust mass over time as they do not contain sufficiently strong pressure bumps to prevent radial inward drift \citep{Birnstiel2010,Pinilla2012a}. \citet{Pinilla2020} demonstrated that disks without pressure bumps drop in dust mass by an factor of 3 in 5 Myr, whereas disks with pressure bumps generally maintain their dust mass. The latter is only true for disks around high mass stars as enhanced dust growth to boulders in disks around low mass stars also decreases the observed dust mass. By including the stellar mass dependence of the presence of substructure (pressure bumps), these models are consistent with the observed trends between dust mass and stellar mass in younger and older regions \citep[Figure 7 in ][]{Pinilla2020} and the two evolutionary pathways observed in Figure \ref{fig:massdistribution} in this work.

Figure \ref{fig:mstarmorphology} demonstrates that compact disks are more common around low-mass stars, which also form the majority of any stellar population: hence the majority of disks drops in dust mass over time, which is a general observation in disk dust mass survey comparisons \citep{Ansdell2017, Cieza2019}. The overall drop in dust mass was initially interpreted as disk dissipation, but this appeared inconsistent with the high accretion rates in the older Upper Sco region \citep{Manara2020}. A modeling approach with both low viscosity (thus slow decrease in accretion rates) and radial drift explains this combined process of disk evolution \citep{Sellek2020}. Our results are compatible with this scenario, with the addition of the structured disks which are unaffected by radial drift. As this is only about 17\% of the total disk population (including the extended disks) and particularly biased towards the highest stellar masses, these structured disks do not have a very pronounced effect on the overall disk evolution: the compact disks dominate the statistics. This scenario is also consistent with the size-luminosity relation derived by \citet{Tripathi2017}, which is overall quadratic consistent with a grain size distribution dominated by radial drift rather than fragmentation \citep{Rosotti2019}, but with larger scatter at larger radii, suggesting that a fraction of the disks is consistent with a grain size distribution set by fragmentation (and thus pressure bumps) instead.
Overall, the presence of giant planets (above the threshold to create pressure bumps) thus sets the evolutionary path of the dust mass and size of the disk.

A second aspect of disk dust mass evolution is that the dust mass decreases faster for lower mass stars (Figure \ref{fig:mdiskmstar}): this is also called the steepening of the $M_{\rm dust}-M_*$ relation \citep{Ansdell2017}. Our results naturally explain this as structured disks are much less common around low-mass stars compared to higher mass stars (Figure \ref{fig:mstarmorphology}), which can be understood immediately as giant planets are also much less common around low-mass stars (Figure \ref{fig:mstarplanets}).  Interestingly, \citet{Pinilla2020} already predicted that strong pressure bumps were required to explain the steepening of the $M_{\rm dust}-M_*$ relation. However, they introduced pressure bumps equally in all stellar mass bins. Strong traps were indeed required for the high stellar masses, but for the low stellar masses the $M_{\rm dust}$ value dropped regardless of the presence of pressure bumps \citep[Figure 7 in][]{Pinilla2020}: for the model with an unperturbed density profile the decrease is due to drift and for the model with pressure bumps due to growth up to boulders. Thus one can conclude that these dust evolution models, when the presence of pressure bumps were introduced with respect to their relative presence in their stellar mass bins, would indeed reproduce our results. 

The dust disk evolution is thus not uniquely set by its stellar mass, but by its capability to form giant planets. As intermediate mass stars have higher mass disks and thus a higher likelihood to form giant planets, they are more likely to have structured disks, so one can still conclude that the majority of the intermediate mass stars are expected to have longer lived disks. Interestingly, \citet{Ribas2015} concluded that disks around intermediate mass stars evolve faster than low mass stars. However, this can be understood as they considered transition disks as evolved disks rather than primordial, as they only considered the infrared fluxes in their work. A drop in infrared flux can be caused by the presence of a large cavity, rather than disk evolution. The millimeter fluxes of transition disks are high and these disks should thus be seen as primordial. 

Another interesting aspect about the disks around intermediate mass stars is the origin of the disks without substructure. Why were these disks incapable of forming giant planets? Inspection of their dust masses reveals that their dust masses are generally somewhat lower than the structured disks. More importantly, more than half of these disks are known wide binaries in the 30-300 au separation range, which might affect the planet formation efficiency in these disks \citep{Harris2012}. Furthermore, we notice that 5 of these disks are so-called Group II disks in Herbig studies \citep{Meeus2001}. In contrast to Group I disks (mostly transition disks), Group II disks are fainter in the infrared and do not show signs of an inner cavity \citep{Maaskant2013}. Their millimeter properties are not well constrained. Based on a large range of properties, \citet{Garufi2017} proposed that Group I and Group II disks evolve through different evolutionary pathways. A similar suggestion was proposed by \citet{vanderMarel2018} for T Tauri stars in Lupus with and without cavities. This suggestion is consistent with our proposed scenario of disk evolution with and without pressure bumps.

\begin{figure}[!ht]
    \centering
    \includegraphics[width=\textwidth]{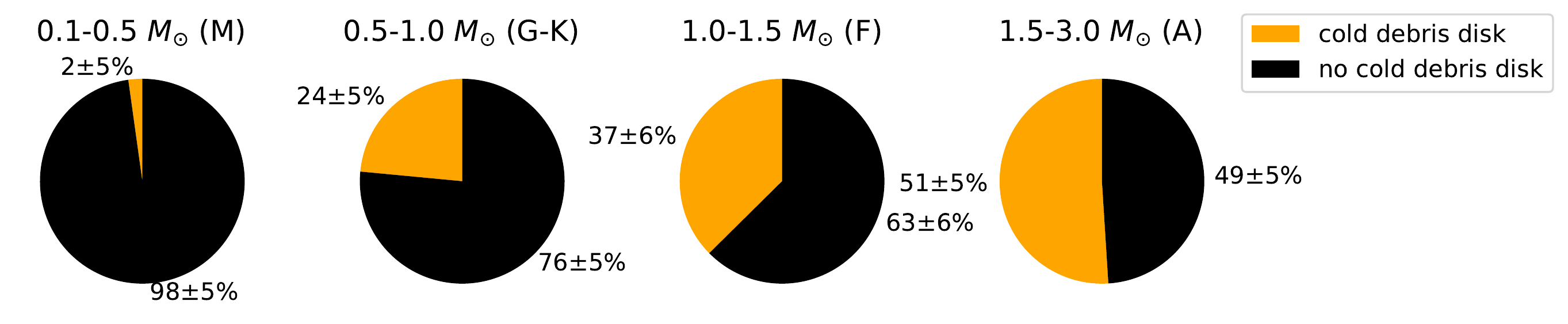}
    \caption{Distribution of cold debris disk fraction across stellar mass bins, following the values derived by \citet{Thureau2014,Sibthorpe2018} based on a completeness analysis of the \textit{Herschel} DEBRIS survey. }
    \label{fig:debrisdisks}
\end{figure}

A final question is what happens at the end of the lifetime of the disk. According to viscous evolution, the disk accretion rate drops with time \citep{Hartmann1998} until eventually photoevaporation takes over and clears the gas in the disk \citep[][and references therein]{Clarke2001,Alexander2014}. At this stage, all pressure bumps will be removed and the dust grains are expected to largely dissipate with the gas. However, the dust that has grown to planetesimal sizes will no longer be affected by the presence or absence of gas, as they are completely decoupled. A ring of planetesimals formed in a pressure bump in the primordial disk would thus remain after most of the disk is dissipated, with a fraction of the original dust mass remaining, dominated by the large solid bodies. Such a planetesimal ring is seen in cold debris disks, where a collisional cascade caused by stirring of the planetesimals results in smaller, observable dust grains \citep[][and references therein]{Wyatt2008,Matthews2014}. Debris disks generally display large dust rings or belts at tens or even hundreds of au \citep{Matra2018}, similar in appearance but much fainter than transition disks. How common are these cold debris disks? According to the \textit{Herschel} DEBRIS survey, the detection rate of debris disks, when corrected for completeness in particular to correct for detection biases towards more luminous stars, increases with stellar mass \citep{Thureau2014,Sibthorpe2018}. In contrast, warm debris disks with dust near the habitable zone show no correlation with
spectral type in LBTI results \citep{Ertel2020}, but structures at these scales in protoplanetary disks cannot be resolved by ALMA.

In Figure \ref{fig:debrisdisks} we plot the cold debris disk fraction for our stellar mass bins. There is again a similarity with the structured disk fractions in Figure \ref{fig:mstarmorphology}, in particular the strong increase with stellar mass, although the fractions do not match exactly. This leads to the hypothesis that the detected cold debris disks are the final outcome of the primordial structured disks, rather than a general representation of protoplanetary disk remnants, as proposed by \citet{Michel2020}. As the majority of protoplanetary disks are affected by radial drift, no cold dust or planetesimals would remain present at large radii. Although stirring is required for the collisional cascade to generate the observable dust debris disks to begin with, it is not unlikely that continuous stirring would occur in a system with at least one giant planet and a surrounding planetesimal belt. This scenario thus suggests that all debris disks must host massive planets, as they are thought to be responsible for clearing the large gaps in the primordial phase. A tentative correlation has indeed been found between the presence of giant planets at wide orbits and bright debris disks \citep{Meshkat2017}, but \citet{Yelverton2020} showed that a correlation between debris disks and close-in planets does not exist. Perhaps the planet mass and detectability in the relevant orbital range can solve this discrepancy, but the connection between debris disks, giant planets and structured disks thus currently remains debatable. 

\subsection{Implications for planet formation}
\label{sct:planetformation}
The proposed scenario suggests that giant planets must form in the first Myr, in order to the create structures that are observed in protoplanetary disks ranging in age from 1-10 Myr old. These planets would then have to migrate inward to 1-10 au, where they are detectable as giant exoplanets around main-sequence stars in radial velocity surveys.
This very early formation is consistent with recent insights into the formation of the solar system. Jupiter must have grown massive enough early on to create a gap that separates different reservoirs of material in the solar nebula with different isotopic composition \citep{Kruijer2017}. 
Jupiter subsequently migrated inward through the disk to $\sim 1.5$ au \citep{Walsh2011}, though the proposed outward migration afterwards may have been unique for the Solar System and does not have to be common for giant exoplanets.

The anti-correlation between Kepler planets and structured protoplanetary disks (Section \ref{sct:kepler}) suggests that the pressure bumps at large disk radii suppress the formation of close-in super-Earths. A plausible mechanisms to achieve this is if these planets form through pebble accretion. \cite{Lambrechts2019} show that the formation of super-Earths is controlled by the magnitude of the flow of pebbles into the inner disk. A reduction in this flow by disk structures would prevent detectable super-Earths from forming. Thus, the high planet occurrence around M dwarfs \citep[e.g.][]{Mulders2015a} could thus be directly connected to the observed lower incidence of structured disks around low-mass stars.  
This scenario also suggests that close-in planetary systems form preferentially in the small, low-mass dust disks that are harder to resolve spatially with ALMA.

Our evolutionary scenario follows along the lines of the separation of super-Earth and terrestrial planet formation pathways as regulated by the pebble flux \citep{Lambrechts2019}, as the pebble flux would be naturally reduced by the presence of giant planets and pressure bumps, leading to formation of terrestrial planets only. Our Solar System does not contain super-Earths but several giant planets and small terrestrial planets, and would thus at some point have been a structured disk with rings and gaps where radial drift and pebble flux is limited, according to this scenario. The large scale structured disks proposed in our study of 40 au and above would be too far out to be explained as limited pebble drift by only Jupiter within this scenario, but Neptune and Uranus are sufficiently far out and would limit pebble drift as well. The Kuiper Belt might be a remnant of the pebbles from a large scale structure disk.

Our planet formation scenario requires a specific timing of the formation of different types of planets: Giant planets form first, super-Earths form later.
This timeline is consistent with the measured properties of Kepler planets. The low density of many of these planets suggest they must have formed while disk gas was still present. In addition, the presence of a radius valley suggest that even planets that are currently rocky ($<1.5 R_\text{Earth}$, \citealt{Rogers2015}) initially formed with a gaseous envelope \citep[e.g.][]{Fulton2017, OwenWu2017,RogersOwen2020} --- and hence, before the gas in the disk had dissipated.
Because sub-Neptunes have accreted only a small fraction of their total mass in gas, this suggests that formation happens towards the end of the disk lifetime when the disk gas is already depleted \citep[e.g.][]{LeeChiang2016,Dawson2016}.

The hypothesis that disk structures induced by giant planets suppress the formation of close-in super-Earths would affect the exoplanet population in multiple ways.
First, it implies that super-Earths and cold giant planets would likely not occur in the same systems. This is slightly at odds with multiple studies that indicates correlations between those planets \citep{Bryan2019,Herman2019} though this is not always found \citep{Barbato2018}. 
This hypothesis also provides an explanation for why the planet occurrence rates of sub-Neptunes follow different scaling relations with stellar mass and metallicity as (cold) giant planets (see \citealt{Mulders2018} for a review on this topic), though we note that binarity may also play a role in explaining the observed (anti)correlations with stellar properties \citep[e.g.][]{MoeKratter2019,KutraWu2020}.

The question remains how giant planets can be present at the large gap radii of tens of au observed in protoplanetary disks, when core accretion is generally not efficient at these radii \citep[e.g.][and references therein]{Helled2014}. One possibility is that planets form quickly in the very inner part of the disk and then initially migrate outwards (Type II migration) to the position they keep during the protoplanetary disk phase \citep{Paardekooper2009}. It is also possible that they form quickly through gravitational instability \citep{Boss1997} in the young, compact circumstellar disk. Planet formation models in the conditions of embedded, compact disks are required to test these scenarios.

\subsection{Uncertainties and biases}
\label{sct:uncertainties}
In this work we have shown that observed fraction of structured disks is positively correlated with stellar mass, that such a correlation matches the occurrence rates of giant exoplanets, and that enough exoplanets of sufficient mass to open disk gaps are available to explain the observed structures. 
However, correlation does not automatically imply causation, and alternative explanations are possible.

One source of possible bias in our study is the ability to detect substructure with ALMA in disks of different size and brightness (hence dust mass). In this work only large scale substructure ($>$25 au) are considered as a tracer of pressure bumps caused by giant planets. 
The concentration of observed transition and ringed disks at high mass and radius may be a result of the preference of high-resolution observation ALMA studies of the brightest and most extended disks, where gaps are more easily resolved. Disk gaps in disks that are smaller or fainter may fall below the detection limits. If these detection biases are indeed significant, the large fraction of transition disks around high mass stars is not a result of those disks having giant planets, but of those disks being larger (and/or more massive). In that scenario all disks may have substructure, but only the large scale substructure leading to larger disk sizes is identified here to lead to a classification as structured disk.  

Disk size is known to be strongly correlated with dust mass \citep{Tripathi2017} with no strong residual dependence on stellar mass \citep{Andrews2018,Hendler2020}. Thus, in this scenario, transition disks would appear more frequently in ALMA observations of more massive disks because these disks are also larger. Giant planets would also preferentially from in these more massive disks, but would not have to be the direct cause of the observed gaps. For example, if giant exoplanets form closer to their current locations, any gaps they cause would be located interior to the resolution of ALMA, and thus not be detected. In particular, our threshold of 40 au disk size to label a disk as 'Extended' (which is added to the ring disk fraction in the comparison with exoplanets) is chosen from the known transition and ring disks and thus possibly caused by this detectability bias. A lower threshold value would result in a larger fraction of extended disks and thus a larger range of planets capable of opening a gap within our interpretation. Furthermore, if giant planets were present in compact, low-mass disks at smaller orbital radii (either through formation or migration), their gaps would remain undetected in our classification. In principle, the fraction of ring/extended disks should be considered a lower limit, which means that the minimum planet mass responsible would be an upper limit. 
 
On the other hand, the Taurus disk survey by \citet{Long2019} was relatively unbiased and does show a clear distinction between compact disks without and extended disks with substructure around 40 au, suggesting that there is a true separation between these two disk populations. Furthermore, if a significant number of compact disks were hiding small scale substructure caused by giant planets, there are insufficient giant planets in the current exoplanet population, in particular around low-mass stars. The stellar mass dependence of the giant exoplanets is thus an argument against this possibility. We emphasize that this conclusion relies on the assumption that all dust substructures are caused by giant planets and that the occurrence rates of giant planets (in particular around low-mass stars) are not underestimated.

On a final note, the extended disks are considered exclusively as ring disks in our analysis, whereas some of these disks might be transitional as well, which poses some additional uncertainty on the derived transition disk fraction, but considering the number of extended disks this is a marginal effect.

A scenario where giant planets preferentially form from the observed substructure is also possible. This is a chicken and egg situation that can only be resolved by directly observing forming planets in disks. The detection of multiple planets embedded in disks with substructure \citep[e.g.][]{Haffert2019,Pinte2019} provides a promising avenue to understand if the statistical relations between (exo)planets and disks identified in this paper are also borne out at the level of individual protoplanetary disks.

\newpage
\subsection{Binarity}
\label{sct:binaries}
Binarity may play an important role in the comparison of disks and exoplanets as well, as binaries can either prevent the formation of a disk, limit its lifetime or truncate the millimeter dust disk either from the outside or inside, i.e. a circumbinary disk \citep{Harris2012}. Binarity is not well constrained for young disk surveys and its influence remains difficult to ascertain. We will just summarize the main statistics of binaries in this section and comment on its role within this study, in particular the role of cirumbinary disks in explaining transition disks. For example, the HD142527 transition disk was found to host a substellar companion \citep{Biller2012,Close2014}. However, for several transition disks stringent upper limits on possible companions exist, excluding this as the main mechanism \citep{Andrews2011,vanderMarel2020}. However, some transition disks in our sample may still be explained by circumbinarity. Rather than studying individual limits we approach this from a statistical perspective of the occurrence rate of binarity as function of stellar mass.

\begin{figure}[!ht]
    \centering
    \includegraphics[width=\textwidth]{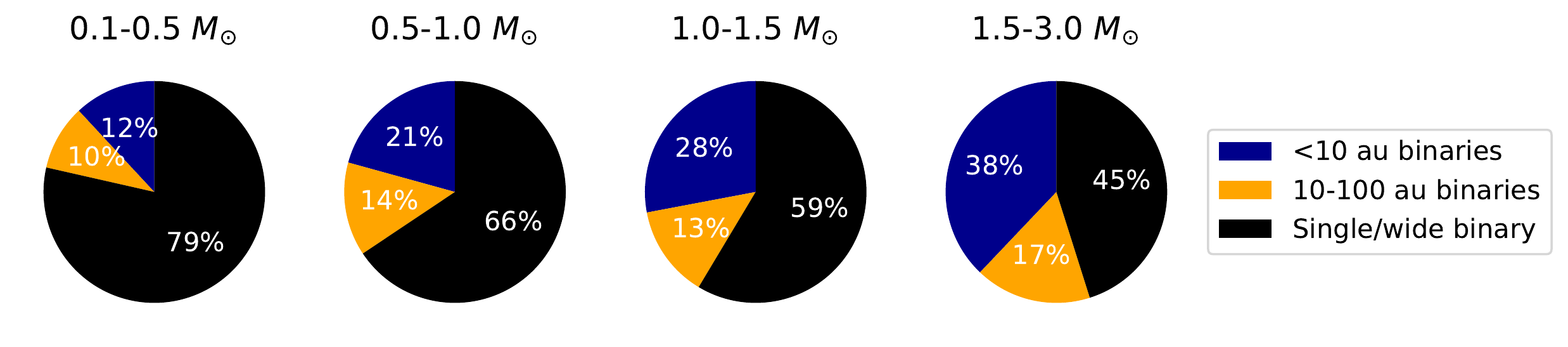}
    \caption{Distribution of binary fraction for different stellar mass bins, based on \citet{MoeKratter2019}. Binaries are separated in close binaries ($<$10 au separation) and gap carving binaries (10-100 au separation).}
    \label{fig:binarity}
\end{figure}

Binarity of stars is known to increase with stellar mass \citep{Raghavan2010} and the binarity fraction has been computed explicitly for two different separation regimes: close ($<$10 au) and wide ($<$100 au) separation by \citet[][Figure 1]{MoeKratter2019} for companions down to 80 $M_{\rm Jup}$ (brown dwarf limit). Using these curves, we compute the fraction of close $<$10 au binaries and wide binaries ($10-100$ au) for each stellar mass bin. The fraction of very wide binaries ($>$100 au) is not quantified, but such wide separations are unlikely to influence the disk structure \citep{Harris2012}. Figure \ref{fig:binarity} shows these binarity fractions. 

Close binarity increases by more than a factor 3 from the lowest to the highest stellar mass bin, but the fraction of 10-100 au separation binaries increases by less than a factor 2. We note that the occurrence rate of brown dwarfs (13-80 $M_{\rm Jup}$) at 10-100 au is much lower than stellar and Super-Jovian companions (brown-dwarf desert): its occurrence is estimated as $\sim$0.8\% for all stars \citep{Nielsen2019}, but as this is not separated for different stellar mass bins, this fraction cannot be included in Figure \ref{fig:binarity}.

Although the binary separation range is somewhat larger than the expected range for clearing transition disk cavities, these fractions may provide a possible explanation for the low occurrence rate of Super-Jovians compared to the number of transition disks, as the fraction of wide stellar companions (for all stars) is also approximately 15\%. An exact comparison is not possible due to the different ranges in separation and the lack of proper statistics on binaries truncating the outer rather than the inner disk \citep{Harris2012}. Also, it is unclear why the overabundance of transition disks is not seen in the lower stellar mass bins, where binarity is decreased, but not down to zero. Likely the disk mass and size distribution with respect to the stellar mass play a role here: if disks around low mass stars are smaller, binaries at 10-100 au separation are less likely to truncate them internally rather than externally. 

Binarity may also play a role in the disk fraction: the number of young stars that actually show an indication of a disk. The infrared disk fraction is known to drop with age \citep{Hernandez2007,Mamajek2009}, although perhaps not as quickly as previously estimated \citep{Michel2020}. ALMA disk surveys are biased as they have pre-selected young stars with infrared excess, i.e. a 100\% disk fraction, whereas even the youngest star forming regions have disk fractions of only 60-80\% which drops to 20-40\% in the older regions. This deficit in young star forming regions might be explained by multiplicity, as the tidal influence of close stellar companions within $\sim$40 au either prevent the formation or limit the lifetime of a disk \citep{Kraus2012}. This implies that the ALMA disk surveys are intrinsically biased towards single stars and binary stars with larger separations. 

On the other hand, exoplanets surveys often have a similar bias towards single stars as known binaries are excluded from the radial velocity target star samples. In transit surveys such as Kepler, however, the binarity of target stars is not always known but can be statistically corrected for \citep[e.g.][]{MoeKratter2019}. In our work we have assumed an average disk fraction of 50\% in the comparison with exoplanet demographics, but it is possible that this is an underestimate due to the bias in each sample. In that case, the number of gap-opening planets in disks would be a factor 2 lower, which could be corrected for by adjusting the minimum planet masses downward by a factor of $\sqrt{2}$.

\section{Conclusions}
We have conducted a large sample study of protoplanetary disks in nearby star forming regions using continuum data from the Atacama Large Millimeter Array. By characterizing the spatial distribution of dust around $\sim 700$ stars ranging in mass from 0.1-3 $M_{\odot}$ considering the effects of dust evolution, and comparing their properties to the population of observed exoplanets, we aim to understand the role that disk gaps and cavities plays in the disk evolution process.
Disks are classified as either transition disk ($>$25 au radius dust cavity), ring disk (one or more dust gaps $<$25 au in radial width), extended disk (no resolved structures but larger than 40 au, implying the disk must have gaps as well) or compact disk (no resolved structures and smaller than 40 au). Transition, ring and extended disks are considered structured, from the arguments of dust evolution, that dust can only be present at large orbital radii when the disk contains pressure bumps to halt the radial drift, under the assumption that dust drift and trapping regulate the disk dust size and morphology. The disk sample and its classifications are compared with age and stellar mass. 
Subsequently, we compare the stellar mass dependence of structured disk occurrence rate with the occurrence rates of exoplanets from radial velocity and transit surveys to understand what type of planets could give rise to the observed disk morphologies. 
This leads to the following conclusions:

\begin{enumerate}
    \item 
    Structure is primarily observed in disks with a high dust mass of $\gtrsim 10 \,M_{\rm Earth}$, compared to the typical dust mass of 0.1-10 $M_{\rm Earth}$ in the sample.
    \item 
    Structure is found in disks with a high dust mass at all ages,  whereas the dust mass of the majority of the disks in the sample decreases with age.
    This can be understood as structured disks contain pressure bumps that prevent the loss of mm-dust material through radial inward drift, whereas the majority of disks without such pressure bumps are more susceptible to loss of mm-dust material via inward radial drift.
    \item The fraction of structured disks, in particular transition disks, is strongly dependent on stellar mass: 
    In particular around the most massive stars in the sample (1.5-3 $M_{\odot}$),
    at least two thirds of the disks is structured. We derive a statistically significant stellar mass power-law dependence of both the transition and ring/extended disk fractions.
    \item Giant exoplanets around main-sequence stars occur with the same frequency as structured disks and display the same stellar mass dependence. This suggests that the observed disk structures could be linked to forming giant planets, provided that they migrate inwards during the lifetime of the disk to the locations where they are currently observed as exoplanets in radial velocity surveys.
    \item A detailed comparison between the observed fraction of structured disks and a demographical model of exoplanets show that:
    1) a minimum planet mass threshold of $1.2\cdot{M_*}^{-1.4} M_{\rm Jup}$ matches the observed fraction of transition disks and 
    2) a minimum threshold of $0.2\cdot{M_*}^{0.1} M_{\rm Jup}$ matches the fraction of ringed/extended disks. 
    This means that transition disk cavities are likely carved by (Super-)Jovian planets and gaps in ring disks by (Super-)Neptune mass planets, consistent with the minimum mass requirement within a factor of a few for forming dust-trapping pressure bumps from planet-disk interaction simulations.
    \item On the other hand, 
    the occurrence rate of transiting planets smaller than Neptune is anti-correlated with stellar mass. This can be understood if those planetary systems form only in disks \emph{without} substructure, where dust can drift unimpeded into the inner disk to form super-Earths via pebble accretion.
    \item These results are consistent with an  evolutionary scenario where the initial gas disk mass determines whether giant planets can form in the disk, which subsequently determines whether pressure bumps are created, resulting in a long-lived structured millimeter-dust disk.
    \item This scenario also predicts that giant planets must form early enough to explain the presence of observed structures in young 1 Myr old disks,
    whereas super-Earths would form later if radial drift is not halted by giant planets that formed farther out.
    \item Stellar binarity may play a moderate role as well in the disk demographics, both by inner truncation of circumbinary disks, resulting in additional transition disks, and through the inhibition of disk formation.
\end{enumerate}

Although our results depend on several assumptions and a non-uniform sample of disk studies with a range of spatial resolutions and sensitivities, the main trends with stellar mass appear to be robust. Future high-resolution disk surveys of compact disks are required to fully confirm the proposed scenario, as well as the direct detection of forming planets in structured disks.

  \begin{acknowledgements}
  \emph{Acknowledgements.} The authors thank the referee for a critical review of our work, which has greatly improved the manuscript. The authors would like to thank Jonathan Williams, Fred Ciesla, Ralph Pudritz, Brenda Matthews and Henry Ngo for useful discussions. We would also like to thank Megan Ansdell for providing the Lupus Band 6 continuum images. N.M. acknowledges support from the Banting Postdoctoral Fellowships program, administered by the Government of Canada. 
  G.D.M. acknowledges support from ANID --- Millennium Science Initiative ---  ICN12\_009.
This paper makes use of several published ALMA datasets (see Table \ref{tbl:regions} for references). ALMA is a partnership of ESO (representing its member states), NSF (USA) and  NINS (Japan), together with NRC (Canada) and NSC and ASIAA (Taiwan), in cooperation with the Republic of Chile. The Joint ALMA Observatory is operated by ESO, AUI/NRAO and NAOJ.  \
 This material is based upon work supported by the National Aeronautics and Space Administration under Agreement No. 80NSSC21K0593 for the program ``Alien Earths''. The results reported herein benefitted from collaborations and/or information exchange within NASA's Nexus for Exoplanet System Science (NExSS) research coordination network sponsored by NASA’s Science Mission Directorate.
  \end{acknowledgements}

\bibliographystyle{aasjournal}

\begin{thebibliography}{}
\expandafter\ifx\csname natexlab\endcsname\relax\def\natexlab#1{#1}\fi
\providecommand{\url}[1]{\href{#1}{#1}}

\bibitem[{{Akeson} \& {Jensen}(2014)}]{Akeson2014}
{Akeson}, R.~L., \& {Jensen}, E.~L.~N. 2014, \apj, 784, 62

\bibitem[{{Akeson} {et~al.}(2019){Akeson}, {Jensen}, {Carpenter}, {Ricci},
  {Laos}, {Nogueira}, \& {Suen-Lewis}}]{Akeson2019}
{Akeson}, R.~L., {Jensen}, E. L.~N., {Carpenter}, J., {et~al.} 2019, \apj, 872,
  158

\bibitem[{{Alcal{\'a}} {et~al.}(2017){Alcal{\'a}}, {Manara}, {Natta}, {Frasca},
  {Testi}, {Nisini}, {Stelzer}, {Williams}, {Antoniucci}, {Biazzo}, {Covino},
  {Esposito}, {Getman}, \& {Rigliaco}}]{Alcala2017}
{Alcal{\'a}}, J.~M., {Manara}, C.~F., {Natta}, A., {et~al.} 2017, \aap, 600,
  A20

\bibitem[{{Alexander} {et~al.}(2014){Alexander}, {Pascucci}, {Andrews},
  {Armitage}, \& {Cieza}}]{Alexander2014}
{Alexander}, R., {Pascucci}, I., {Andrews}, S., {Armitage}, P., \& {Cieza}, L.
  2014, in Protostars and Planets VI, ed. H.~{Beuther}, R.~S. {Klessen}, C.~P.
  {Dullemond}, \& T.~{Henning}, 475

\bibitem[{{ALMA Partnership} {et~al.}(2015){ALMA Partnership}, {Brogan},
  {P{\'e}rez}, {Hunter}, {Dent}, {Hales}, {Hills}, {Corder}, {Fomalont},
  {Vlahakis}, {Asaki}, {Barkats}, {Hirota}, {Hodge}, {Impellizzeri}, {Kneissl},
  {Liuzzo}, {Lucas}, {Marcelino}, {Matsushita}, {Nakanishi}, {Phillips},
  {Richards}, {Toledo}, {Aladro}, {Broguiere}, {Cortes}, {Cortes}, {Espada},
  {Galarza}, {Garcia-Appadoo}, {Guzman-Ramirez}, {Humphreys}, {Jung}, {Kameno},
  {Laing}, {Leon}, {Marconi}, {Mignano}, {Nikolic}, {Nyman}, {Radiszcz},
  {Remijan}, {Rod{\'o}n}, {Sawada}, {Takahashi}, {Tilanus}, {Vila Vilaro},
  {Watson}, {Wiklind}, {Akiyama}, {Chapillon}, {de Gregorio-Monsalvo}, {Di
  Francesco}, {Gueth}, {Kawamura}, {Lee}, {Nguyen Luong}, {Mangum}, {Pietu},
  {Sanhueza}, {Saigo}, {Takakuwa}, {Ubach}, {van Kempen}, {Wootten},
  {Castro-Carrizo}, {Francke}, {Gallardo}, {Garcia}, {Gonzalez}, {Hill},
  {Kaminski}, {Kurono}, {Liu}, {Lopez}, {Morales}, {Plarre}, {Schieven},
  {Testi}, {Videla}, {Villard}, {Andreani}, {Hibbard}, \&
  {Tatematsu}}]{HLTau2015}
{ALMA Partnership}, A., {Brogan}, C.~L., {P{\'e}rez}, L.~M., {et~al.} 2015,
  \apjl, 808, L3

\bibitem[{{Andrews}(2020)}]{Andrews2020}
{Andrews}, S.~M. 2020, arXiv e-prints, arXiv:2001.05007

\bibitem[{{Andrews} {et~al.}(2010{\natexlab{a}}){Andrews}, {Czekala}, {Wilner},
  {Espaillat}, {Dullemond}, \& {Hughes}}]{Andrews2010}
{Andrews}, S.~M., {Czekala}, I., {Wilner}, D.~J., {et~al.} 2010{\natexlab{a}},
  \apj, 710, 462

\bibitem[{{Andrews} {et~al.}(2010{\natexlab{b}}){Andrews}, {Czekala}, {Wilner},
  {Espaillat}, {Dullemond}, \& {Hughes}}]{Andrews2009}
---. 2010{\natexlab{b}}, \apj, 710, 462

\bibitem[{{Andrews} {et~al.}(2013){Andrews}, {Rosenfeld}, {Kraus}, \&
  {Wilner}}]{Andrews2013}
{Andrews}, S.~M., {Rosenfeld}, K.~A., {Kraus}, A.~L., \& {Wilner}, D.~J. 2013,
  \apj, 771, 129

\bibitem[{{Andrews} {et~al.}(2018){Andrews}, {Terrell}, {Tripathi}, {Ansdell},
  {Williams}, \& {Wilner}}]{Andrews2018}
{Andrews}, S.~M., {Terrell}, M., {Tripathi}, A., {et~al.} 2018, \apj, 865, 157

\bibitem[{{Andrews} {et~al.}(2011){Andrews}, {Wilner}, {Espaillat}, {Hughes},
  {Dullemond}, {McClure}, {Qi}, \& {Brown}}]{Andrews2011}
{Andrews}, S.~M., {Wilner}, D.~J., {Espaillat}, C., {et~al.} 2011, \apj, 732,
  42

\bibitem[{{Andrews} {et~al.}(2016){Andrews}, {Wilner}, {Zhu}, {Birnstiel},
  {Carpenter}, {P{\'e}rez}, {Bai}, {{\"O}berg}, {Hughes}, {Isella}, \&
  {Ricci}}]{Andrews2016}
{Andrews}, S.~M., {Wilner}, D.~J., {Zhu}, Z., {et~al.} 2016, \apjl, 820, L40

\bibitem[{{Ansdell} {et~al.}(2017){Ansdell}, {Williams}, {Manara}, {Miotello},
  {Facchini}, {van der Marel}, {Testi}, \& {van Dishoeck}}]{Ansdell2017}
{Ansdell}, M., {Williams}, J.~P., {Manara}, C.~F., {et~al.} 2017, \aj, 153, 240

\bibitem[{{Ansdell} {et~al.}(2016){Ansdell}, {Williams}, {van der Marel},
  {Carpenter}, {Guidi}, {Hogerheijde}, {Mathews}, {Manara}, {Miotello},
  {Natta}, {Oliveira}, {Tazzari}, {Testi}, {van Dishoeck}, \& {van
  Terwisga}}]{Ansdell2016}
{Ansdell}, M., {Williams}, J.~P., {van der Marel}, N., {et~al.} 2016, \apj,
  828, 46

\bibitem[{{Ansdell} {et~al.}(2018){Ansdell}, {Williams}, {Trapman}, {van
  Terwisga}, {Facchini}, {Manara}, {van der Marel}, {Miotello}, {Tazzari},
  {Hogerheijde}, {Guidi}, {Testi}, \& {van Dishoeck}}]{Ansdell2018}
{Ansdell}, M., {Williams}, J.~P., {Trapman}, L., {et~al.} 2018, \apj, 859, 21

\bibitem[{{Ansdell} {et~al.}(2020){Ansdell}, {Haworth}, {Williams}, {Facchini},
  {Winter}, {Manara}, {Hacar}, {Chiang}, {van Terwisga}, {van der Marel}, \&
  {van Dishoeck}}]{Ansdell2020}
{Ansdell}, M., {Haworth}, T.~J., {Williams}, J.~P., {et~al.} 2020, \aj, 160,
  248

\bibitem[{{Artymowicz} \& {Lubow}(1994)}]{Artymowicz1994}
{Artymowicz}, P., \& {Lubow}, S.~H. 1994, \apj, 421, 651

\bibitem[{{Bacciotti} {et~al.}(2018){Bacciotti}, {Girart}, {Padovani}, {Podio},
  {Paladino}, {Testi}, {Bianchi}, {Galli}, {Codella}, {Coffey}, {Favre}, \&
  {Fedele}}]{Bacciotti2018}
{Bacciotti}, F., {Girart}, J.~M., {Padovani}, M., {et~al.} 2018, \apjl, 865,
  L12

\bibitem[{{Barbato} {et~al.}(2018){Barbato}, {Sozzetti}, {Desidera}, {Damasso},
  {Bonomo}, {Giacobbe}, {Colombo}, {Lazzoni}, {Claudi}, {Gratton}, {LoCurto},
  {Marzari}, \& {Mordasini}}]{Barbato2018}
{Barbato}, D., {Sozzetti}, A., {Desidera}, S., {et~al.} 2018, \aap, 615, A175

\bibitem[{{Barenfeld} {et~al.}(2016){Barenfeld}, {Carpenter}, {Ricci}, \&
  {Isella}}]{Barenfeld2016}
{Barenfeld}, S.~A., {Carpenter}, J.~M., {Ricci}, L., \& {Isella}, A. 2016,
  \apj, 827, 142

\bibitem[{{Bell} {et~al.}(2015){Bell}, {Mamajek}, \& {Naylor}}]{Bell2015}
{Bell}, C. P.~M., {Mamajek}, E.~E., \& {Naylor}, T. 2015, \mnras, 454, 593

\bibitem[{{Biller} {et~al.}(2012){Biller}, {Lacour}, {Juh{\'a}sz}, {Benisty},
  {Chauvin}, {Olofsson}, {Pott}, {M{\"u}ller}, {Sicilia-Aguilar}, {Bonnefoy},
  {Tuthill}, {Thebault}, {Henning}, \& {Crida}}]{Biller2012}
{Biller}, B., {Lacour}, S., {Juh{\'a}sz}, A., {et~al.} 2012, \apjl, 753, L38

\bibitem[{{Birnstiel} {et~al.}(2010){Birnstiel}, {Dullemond}, \&
  {Brauer}}]{Birnstiel2010}
{Birnstiel}, T., {Dullemond}, C.~P., \& {Brauer}, F. 2010, \aap, 513, A79

\bibitem[{{Boss}(1997)}]{Boss1997}
{Boss}, A.~P. 1997, Science, 276, 1836

\bibitem[{{Brown} {et~al.}(2009){Brown}, {Blake}, {Qi}, {Dullemond}, {Wilner},
  \& {Williams}}]{Brown2009}
{Brown}, J.~M., {Blake}, G.~A., {Qi}, C., {et~al.} 2009, \apj, 704, 496

\bibitem[{{Bryan} {et~al.}(2019){Bryan}, {Knutson}, {Lee}, {Fulton}, {Batygin},
  {Ngo}, \& {Meshkat}}]{Bryan2019}
{Bryan}, M.~L., {Knutson}, H.~A., {Lee}, E.~J., {et~al.} 2019, \aj, 157, 52

\bibitem[{{Burke} \& {Catanzarite}(2017)}]{Burke2017}
{Burke}, C.~J., \& {Catanzarite}, J. 2017, {Planet Detection Metrics:
  Per-Target Detection Contours for Data Release 25}, Tech. rep.

\bibitem[{{Casassus} {et~al.}(2013){Casassus}, {van der Plas}, {Perez}, {Dent},
  {Fomalont}, {Hagelberg}, {Hales}, {Jord{\'a}n}, {Mawet}, {M{\'e}nard},
  {Wootten}, {Wilner}, {Hughes}, {Schreiber}, {Girard}, {Ercolano}, {Canovas},
  {Rom{\'a}n}, \& {Salinas}}]{Casassus2013}
{Casassus}, S., {van der Plas}, G.~M., {Perez}, S., {et~al.} 2013, \nat, 493,
  191

\bibitem[{{Cazzoletti} {et~al.}(2019){Cazzoletti}, {Manara}, {Baobab Liu}, {van
  Dishoeck}, {Facchini}, {Alcal{\`a}}, {Ansdell}, {Testi}, {Williams},
  {Carrasco-Gonz{\'a}lez}, {Dong}, {Forbrich}, {Fukagawa}, {Galv{\'a}n-Madrid},
  {Hirano}, {Hogerheijde}, {Hasegawa}, {Muto}, {Pinilla}, {Takami}, {Tamura},
  {Tazzari}, \& {Wisniewski}}]{Cazzoletti2019}
{Cazzoletti}, P., {Manara}, C.~F., {Baobab Liu}, H., {et~al.} 2019, \aap, 626,
  A11

\bibitem[{{Chen} {et~al.}(2011){Chen}, {Mamajek}, {Bitner}, {Pecaut}, {Su}, \&
  {Weinberger}}]{Chen2011}
{Chen}, C.~H., {Mamajek}, E.~E., {Bitner}, M.~A., {et~al.} 2011, \apj, 738, 122

\bibitem[{{Chen} {et~al.}(2012){Chen}, {Pecaut}, {Mamajek}, {Su}, \&
  {Bitner}}]{Chen2012}
{Chen}, C.~H., {Pecaut}, M., {Mamajek}, E.~E., {Su}, K. Y.~L., \& {Bitner}, M.
  2012, \apj, 756, 133

\bibitem[{{Cieza} {et~al.}(2019){Cieza}, {Ru{\'\i}z-Rodr{\'\i}guez}, {Hales},
  {Casassus}, {P{\'e}rez}, {Gonzalez-Ruilova}, {C{\'a}novas}, {Williams},
  {Zurlo}, {Ansdell}, {Avenhaus}, {Bayo}, {Bertrang}, {Christiaens}, {Dent},
  {Ferrero}, {Gamen}, {Olofsson}, {Orcajo}, {Pe{\~n}a Ram{\'\i}rez},
  {Principe}, {Schreiber}, \& {van der Plas}}]{Cieza2019}
{Cieza}, L.~A., {Ru{\'\i}z-Rodr{\'\i}guez}, D., {Hales}, A., {et~al.} 2019,
  \mnras, 482, 698

\bibitem[{{Clarke} {et~al.}(2001){Clarke}, {Gendrin}, \&
  {Sotomayor}}]{Clarke2001}
{Clarke}, C.~J., {Gendrin}, A., \& {Sotomayor}, M. 2001, \mnras, 328, 485

\bibitem[{{Close}(2020)}]{Close2020}
{Close}, L.~M. 2020, \aj, 160, 221

\bibitem[{{Close} {et~al.}(2014){Close}, {Follette}, {Males}, {Puglisi},
  {Xompero}, {Apai}, {Najita}, {Weinberger}, {Morzinski}, {Rodigas}, {Hinz},
  {Bailey}, \& {Briguglio}}]{Close2014}
{Close}, L.~M., {Follette}, K.~B., {Males}, J.~R., {et~al.} 2014, \apjl, 781,
  L30

\bibitem[{{Comer{\'o}n}(2008)}]{Comeron2008}
{Comer{\'o}n}, F. 2008, {The Lupus Clouds}, ed. B.~{Reipurth}, 295

\bibitem[{{Cumming} {et~al.}(2008){Cumming}, {Butler}, {Marcy}, {Vogt},
  {Wright}, \& {Fischer}}]{Cumming2008}
{Cumming}, A., {Butler}, R.~P., {Marcy}, G.~W., {et~al.} 2008, \pasp, 120, 531

\bibitem[{{D'Angelo} {et~al.}(2006){D'Angelo}, {Lubow}, \&
  {Bate}}]{dAngelo2006}
{D'Angelo}, G., {Lubow}, S.~H., \& {Bate}, M.~R. 2006, \apj, 652, 1698

\bibitem[{{Dawson} {et~al.}(2016){Dawson}, {Lee}, \& {Chiang}}]{Dawson2016}
{Dawson}, R.~I., {Lee}, E.~J., \& {Chiang}, E. 2016, \apj, 822, 54

\bibitem[{{Dodson-Robinson} \& {Salyk}(2011)}]{Dodson-Robinson2011}
{Dodson-Robinson}, S.~E., \& {Salyk}, C. 2011, \apj, 738, 131

\bibitem[{{Dong} \& {Zhu}(2013)}]{DongZhu2013}
{Dong}, S., \& {Zhu}, Z. 2013, \apj, 778, 53

\bibitem[{{Ertel} {et~al.}(2020){Ertel}, {Defr{\`e}re}, {Hinz}, {Mennesson},
  {Kennedy}, {Danchi}, {Gelino}, {Hill}, {Hoffmann}, {Mazoyer}, {Rieke},
  {Shannon}, {Stapelfeldt}, {Spalding}, {Stone}, {Vaz}, {Weinberger},
  {Willems}, {Absil}, {Arbo}, {Bailey}, {Beichman}, {Bryden}, {Downey},
  {Durney}, {Esposito}, {Gaspar}, {Grenz}, {Haniff}, {Leisenring}, {Marion},
  {McMahon}, {Millan-Gabet}, {Montoya}, {Morzinski}, {Perera}, {Pinna}, {Pott},
  {Power}, {Puglisi}, {Roberge}, {Serabyn}, {Skemer}, {Su}, {Vaitheeswaran}, \&
  {Wyatt}}]{Ertel2020}
{Ertel}, S., {Defr{\`e}re}, D., {Hinz}, P., {et~al.} 2020, \aj, 159, 177

\bibitem[{{Espaillat} {et~al.}(2014){Espaillat}, {Muzerolle}, {Najita},
  {Andrews}, {Zhu}, {Calvet}, {Kraus}, {Hashimoto}, {Kraus}, \&
  {D'Alessio}}]{Espaillat2014}
{Espaillat}, C., {Muzerolle}, J., {Najita}, J., {et~al.} 2014, Protostars and
  Planets VI, 497

\bibitem[{{Esplin} {et~al.}(2018){Esplin}, {Luhman}, {Miller}, \&
  {Mamajek}}]{Esplin2018}
{Esplin}, T.~L., {Luhman}, K.~L., {Miller}, E.~B., \& {Mamajek}, E.~E. 2018,
  \aj, 156, 75

\bibitem[{{Facchini} {et~al.}(2019){Facchini}, {van Dishoeck}, {Manara},
  {Tazzari}, {Maud}, {Cazzoletti}, {Rosotti}, {van der Marel}, {Pinilla}, \&
  {Clarke}}]{Facchini2019}
{Facchini}, S., {van Dishoeck}, E.~F., {Manara}, C.~F., {et~al.} 2019, \aap,
  626, L2

\bibitem[{{Fedele} {et~al.}(2017){Fedele}, {Carney}, {Hogerheijde}, {Walsh},
  {Miotello}, {Klaassen}, {Bruderer}, {Henning}, \& {van
  Dishoeck}}]{Fedele2017}
{Fedele}, D., {Carney}, M., {Hogerheijde}, M.~R., {et~al.} 2017, \aap, 600, A72

\bibitem[{{Fernandes} {et~al.}(2019){Fernandes}, {Mulders}, {Pascucci},
  {Mordasini}, \& {Emsenhuber}}]{Fernandes2019}
{Fernandes}, R.~B., {Mulders}, G.~D., {Pascucci}, I., {Mordasini}, C., \&
  {Emsenhuber}, A. 2019, \apj, 874, 81

\bibitem[{{Foreman-Mackey} {et~al.}(2013){Foreman-Mackey}, {Hogg}, {Lang}, \&
  {Goodman}}]{DFM2013}
{Foreman-Mackey}, D., {Hogg}, D.~W., {Lang}, D., \& {Goodman}, J. 2013, \pasp,
  125, 306

\bibitem[{{Francis} \& {van der Marel}(2020)}]{Francis2020}
{Francis}, L., \& {van der Marel}, N. 2020, \apj, 892, 111

\bibitem[{{Fulton} {et~al.}(2017){Fulton}, {Petigura}, {Howard}, {Isaacson},
  {Marcy}, {Cargile}, {Hebb}, {Weiss}, {Johnson}, {Morton}, {Sinukoff},
  {Crossfield}, \& {Hirsch}}]{Fulton2017}
{Fulton}, B.~J., {Petigura}, E.~A., {Howard}, A.~W., {et~al.} 2017, \aj, 154,
  109

\bibitem[{{Garufi} {et~al.}(2017){Garufi}, {Meeus}, {Benisty}, {Quanz},
  {Banzatti}, {Kama}, {Canovas}, {Eiroa}, {Schmid}, {Stolker}, {Pohl},
  {Rigliaco}, {M{\'e}nard}, {Meyer}, {van Boekel}, \& {Dominik}}]{Garufi2017}
{Garufi}, A., {Meeus}, G., {Benisty}, M., {et~al.} 2017, \aap, 603, A21

\bibitem[{{Garufi} {et~al.}(2018){Garufi}, {Benisty}, {Pinilla}, {Tazzari},
  {Dominik}, {Ginski}, {Henning}, {Kral}, {Langlois}, {M{\'e}nard}, {Stolker},
  {Szulagyi}, {Villenave}, \& {van der Plas}}]{Garufi2018}
{Garufi}, A., {Benisty}, M., {Pinilla}, P., {et~al.} 2018, \aap, 620, A94

\bibitem[{{Ghezzi} {et~al.}(2018){Ghezzi}, {Montet}, \& {Johnson}}]{Ghezzi2018}
{Ghezzi}, L., {Montet}, B.~T., \& {Johnson}, J.~A. 2018, \apj, 860, 109

\bibitem[{{Haffert} {et~al.}(2019){Haffert}, {Bohn}, {de Boer}, {Snellen},
  {Brinchmann}, {Girard}, {Keller}, \& {Bacon}}]{Haffert2019}
{Haffert}, S.~Y., {Bohn}, A.~J., {de Boer}, J., {et~al.} 2019, Nature
  Astronomy, 3, 749

\bibitem[{{Hardy} {et~al.}(2015){Hardy}, {Caceres}, {Schreiber}, {Cieza},
  {Alexander}, {Canovas}, {Williams}, {Wahhaj}, \& {Menard}}]{Hardy2015}
{Hardy}, A., {Caceres}, C., {Schreiber}, M.~R., {et~al.} 2015, \aap, 583, A66

\bibitem[{{Harris} {et~al.}(2012){Harris}, {Andrews}, {Wilner}, \&
  {Kraus}}]{Harris2012}
{Harris}, R.~J., {Andrews}, S.~M., {Wilner}, D.~J., \& {Kraus}, A.~L. 2012,
  \apj, 751, 115

\bibitem[{{Hartmann} {et~al.}(1998){Hartmann}, {Calvet}, {Gullbring}, \&
  {D'Alessio}}]{Hartmann1998}
{Hartmann}, L., {Calvet}, N., {Gullbring}, E., \& {D'Alessio}, P. 1998, \apj,
  495, 385

\bibitem[{{He} {et~al.}(2019){He}, {Ford}, \& {Ragozzine}}]{He2019}
{He}, M.~Y., {Ford}, E.~B., \& {Ragozzine}, D. 2019, \mnras, 490, 4575

\bibitem[{{He} {et~al.}(2021){He}, {Ford}, \& {Ragozzine}}]{He2020}
---. 2021, \aj, 161, 16

\bibitem[{{Helled} {et~al.}(2014){Helled}, {Bodenheimer}, {Podolak}, {Boley},
  {Meru}, {Nayakshin}, {Fortney}, {Mayer}, {Alibert}, \& {Boss}}]{Helled2014}
{Helled}, R., {Bodenheimer}, P., {Podolak}, M., {et~al.} 2014, in Protostars
  and Planets VI, ed. H.~{Beuther}, R.~S. {Klessen}, C.~P. {Dullemond}, \&
  T.~{Henning}, 643

\bibitem[{{Hendler} {et~al.}(2020){Hendler}, {Pascucci}, {Pinilla}, {Tazzari},
  {Carpenter}, {Malhotra}, \& {Testi}}]{Hendler2020}
{Hendler}, N., {Pascucci}, I., {Pinilla}, P., {et~al.} 2020, \apj, 895, 126

\bibitem[{{Herman} {et~al.}(2019){Herman}, {Zhu}, \& {Wu}}]{Herman2019}
{Herman}, M.~K., {Zhu}, W., \& {Wu}, Y. 2019, \aj, 157, 248

\bibitem[{{Hern{\'a}ndez} {et~al.}(2007){Hern{\'a}ndez}, {Hartmann}, {Megeath},
  {Gutermuth}, {Muzerolle}, {Calvet}, {Vivas}, {Brice{\~n}o}, {Allen},
  {Stauffer}, {Young}, \& {Fazio}}]{Hernandez2007}
{Hern{\'a}ndez}, J., {Hartmann}, L., {Megeath}, T., {et~al.} 2007, \apj, 662,
  1067

\bibitem[{{Hildebrand}(1983)}]{Hildebrand1983}
{Hildebrand}, R.~H. 1983, \qjras, 24, 267

\bibitem[{{Hirsh} {et~al.}(2020){Hirsh}, {Price}, {Gonzalez},
  {Ubeira-Gabellini}, \& {Ragusa}}]{Hirsh2020}
{Hirsh}, K., {Price}, D.~J., {Gonzalez}, J.-F., {Ubeira-Gabellini}, M.~G., \&
  {Ragusa}, E. 2020, \mnras, 498, 2936

\bibitem[{{Howard} {et~al.}(2010){Howard}, {Marcy}, {Johnson}, {Fischer},
  {Wright}, {Isaacson}, {Valenti}, {Anderson}, {Lin}, \& {Ida}}]{Howard2010}
{Howard}, A.~W., {Marcy}, G.~W., {Johnson}, J.~A., {et~al.} 2010, Science, 330,
  653

\bibitem[{{Howard} {et~al.}(2012){Howard}, {Marcy}, {Bryson}, {Jenkins},
  {Rowe}, {Batalha}, {Borucki}, {Koch}, {Dunham}, {Gautier}, {Van Cleve},
  {Cochran}, {Latham}, {Lissauer}, {Torres}, {Brown}, {Gilliland}, {Buchhave},
  {Caldwell}, {Christensen-Dalsgaard}, {Ciardi}, {Fressin}, {Haas}, {Howell},
  {Kjeldsen}, {Seager}, {Rogers}, {Sasselov}, {Steffen}, {Basri},
  {Charbonneau}, {Christiansen}, {Clarke}, {Dupree}, {Fabrycky}, {Fischer},
  {Ford}, {Fortney}, {Tarter}, {Girouard}, {Holman}, {Johnson}, {Klaus},
  {Machalek}, {Moorhead}, {Morehead}, {Ragozzine}, {Tenenbaum}, {Twicken},
  {Quinn}, {Isaacson}, {Shporer}, {Lucas}, {Walkowicz}, {Welsh}, {Boss},
  {Devore}, {Gould}, {Smith}, {Morris}, {Prsa}, {Morton}, {Still}, {Thompson},
  {Mullally}, {Endl}, \& {MacQueen}}]{Howard2012}
{Howard}, A.~W., {Marcy}, G.~W., {Bryson}, S.~T., {et~al.} 2012, \apjs, 201, 15

\bibitem[{{Hsu} {et~al.}(2019){Hsu}, {Ford}, {Ragozzine}, \& {Ashby}}]{Hsu2019}
{Hsu}, D.~C., {Ford}, E.~B., {Ragozzine}, D., \& {Ashby}, K. 2019, \aj, 158,
  109

\bibitem[{{Huang} {et~al.}(2018){Huang}, {Andrews}, {P{\'e}rez}, {Zhu},
  {Dullemond}, {Isella}, {Benisty}, {Bai}, {Birnstiel}, {Carpenter},
  {Guzm{\'a}n}, {Hughes}, {{\"O}berg}, {Ricci}, {Wilner}, \&
  {Zhang}}]{Huang2018}
{Huang}, J., {Andrews}, S.~M., {P{\'e}rez}, L.~M., {et~al.} 2018, \apjl, 869,
  L43

\bibitem[{{Isella} {et~al.}(2016){Isella}, {Guidi}, {Testi}, {Liu}, {Li}, {Li},
  {Weaver}, {Boehler}, {Carperter}, {De Gregorio-Monsalvo}, {Manara}, {Natta},
  {P{\'e}rez}, {Ricci}, {Sargent}, {Tazzari}, \& {Turner}}]{Isella2016}
{Isella}, A., {Guidi}, G., {Testi}, L., {et~al.} 2016, Physical Review Letters,
  117, 251101

\bibitem[{{Izidoro} {et~al.}(2015){Izidoro}, {Raymond}, {Morbidelli},
  {Hersant}, \& {Pierens}}]{Izidoro2015}
{Izidoro}, A., {Raymond}, S.~N., {Morbidelli}, A., {Hersant}, F., \& {Pierens},
  A. 2015, \apjl, 800, L22

\bibitem[{{Johansen} {et~al.}(2009){Johansen}, {Youdin}, \&
  {Klahr}}]{Johansen2009}
{Johansen}, A., {Youdin}, A., \& {Klahr}, H. 2009, \apj, 697, 1269

\bibitem[{{Johnson} {et~al.}(2010){Johnson}, {Aller}, {Howard}, \&
  {Crepp}}]{Johnson2010}
{Johnson}, J.~A., {Aller}, K.~M., {Howard}, A.~W., \& {Crepp}, J.~R. 2010,
  \pasp, 122, 905

\bibitem[{{Kelly}(2007)}]{Kelly2007}
{Kelly}, B.~C. 2007, \apj, 665, 1489

\bibitem[{{Keppler} {et~al.}(2018){Keppler}, {Benisty}, {M{\"u}ller},
  {Henning}, {van Boekel}, {Cantalloube}, {Ginski}, {van Holstein}, {Maire},
  {Pohl}, {Samland}, {Avenhaus}, {Baudino}, {Boccaletti}, {de Boer},
  {Bonnefoy}, {Chauvin}, {Desidera}, {Langlois}, {Lazzoni}, {Marleau},
  {Mordasini}, {Pawellek}, {Stolker}, {Vigan}, {Zurlo}, {Birnstiel},
  {Brandner}, {Feldt}, {Flock}, {Girard}, {Gratton}, {Hagelberg}, {Isella},
  {Janson}, {Juhasz}, {Kemmer}, {Kral}, {Lagrange}, {Launhardt}, {Matter},
  {M{\'e}nard}, {Milli}, {Molli{\`e}re}, {Olofsson}, {P{\'e}rez}, {Pinilla},
  {Pinte}, {Quanz}, {Schmidt}, {Udry}, {Wahhaj}, {Williams}, {Buenzli},
  {Cudel}, {Dominik}, {Galicher}, {Kasper}, {Lannier}, {Mesa}, {Mouillet},
  {Peretti}, {Perrot}, {Salter}, {Sissa}, {Wildi}, {Abe}, {Antichi},
  {Augereau}, {Baruffolo}, {Baudoz}, {Bazzon}, {Beuzit}, {Blanchard}, {Brems},
  {Buey}, {De Caprio}, {Carbillet}, {Carle}, {Cascone}, {Cheetham}, {Claudi},
  {Costille}, {Delboulb{\'e}}, {Dohlen}, {Fantinel}, {Feautrier}, {Fusco},
  {Giro}, {Gluck}, {Gry}, {Hubin}, {Hugot}, {Jaquet}, {Le Mignant}, {Llored},
  {Madec}, {Magnard}, {Martinez}, {Maurel}, {Meyer}, {M{\"o}ller-Nilsson},
  {Moulin}, {Mugnier}, {Orign{\'e}}, {Pavlov}, {Perret}, {Petit}, {Pragt},
  {Puget}, {Rabou}, {Ramos}, {Rigal}, {Rochat}, {Roelfsema}, {Rousset}, {Roux},
  {Salasnich}, {Sauvage}, {Sevin}, {Soenke}, {Stadler}, {Suarez}, {Turatto}, \&
  {Weber}}]{Keppler2018}
{Keppler}, M., {Benisty}, M., {M{\"u}ller}, A., {et~al.} 2018, \aap, 617, A44

\bibitem[{{Keppler} {et~al.}(2019){Keppler}, {Teague}, {Bae}, {Benisty},
  {Henning}, {van Boekel}, {Chapillon}, {Pinilla}, {Williams}, {Bertrang},
  {Facchini}, {Flock}, {Ginski}, {Juhasz}, {Klahr}, {Liu}, {M{\"u}ller},
  {P{\'e}rez}, {Pohl}, {Rosotti}, {Samland}, \& {Semenov}}]{Keppler2019}
{Keppler}, M., {Teague}, R., {Bae}, J., {et~al.} 2019, \aap, 625, A118

\bibitem[{{Kley} \& {Dirksen}(2006)}]{kleyDirksen2006}
{Kley}, W., \& {Dirksen}, G. 2006, \aap, 447, 369

\bibitem[{{Kley} \& {Nelson}(2012)}]{KleyNelson2012}
{Kley}, W., \& {Nelson}, R.~P. 2012, \araa, 50, 211

\bibitem[{{Kraus} \& {Hillenbrand}(2009)}]{Kraus2009}
{Kraus}, A.~L., \& {Hillenbrand}, L.~A. 2009, \apj, 704, 531

\bibitem[{{Kraus} {et~al.}(2012){Kraus}, {Ireland}, {Hillenbrand}, \&
  {Martinache}}]{Kraus2012}
{Kraus}, A.~L., {Ireland}, M.~J., {Hillenbrand}, L.~A., \& {Martinache}, F.
  2012, \apj, 745, 19

\bibitem[{{Kruijer} {et~al.}(2017){Kruijer}, {Burkhardt}, {Budde}, \&
  {Kleine}}]{Kruijer2017}
{Kruijer}, T.~S., {Burkhardt}, C., {Budde}, G., \& {Kleine}, T. 2017,
  Proceedings of the National Academy of Science, 114, 6712

\bibitem[{{Kutra} \& {Wu}(2020)}]{KutraWu2020}
{Kutra}, T., \& {Wu}, Y. 2020, arXiv e-prints, arXiv:2003.08431

\bibitem[{{Lada} \& {Wilking}(1984)}]{Lada1984}
{Lada}, C.~J., \& {Wilking}, B.~A. 1984, \apj, 287, 610

\bibitem[{{Lambrechts} {et~al.}(2019){Lambrechts}, {Morbidelli}, {Jacobson},
  {Johansen}, {Bitsch}, {Izidoro}, \& {Raymond}}]{Lambrechts2019}
{Lambrechts}, M., {Morbidelli}, A., {Jacobson}, S.~A., {et~al.} 2019, \aap,
  627, A83

\bibitem[{{Lee} \& {Chiang}(2016)}]{LeeChiang2016}
{Lee}, E.~J., \& {Chiang}, E. 2016, \apj, 817, 90

\bibitem[{{Lin} \& {Papaloizou}(1979)}]{LinPapaloizou1979}
{Lin}, D.~N.~C., \& {Papaloizou}, J. 1979, \mnras, 188, 191

\bibitem[{{Lodato} {et~al.}(2019){Lodato}, {Dipierro}, {Ragusa}, {Long},
  {Herczeg}, {Pascucci}, {Pinilla}, {Manara}, {Tazzari}, {Liu}, {Mulders},
  {Harsono}, {Boehler}, {M{\'e}nard}, {Johnstone}, {Salyk}, {van der Plas},
  {Cabrit}, {Edwards}, {Fischer}, {Hendler}, {Nisini}, {Rigliaco}, {Avenhaus},
  {Banzatti}, \& {Gully-Santiago}}]{Lodato2019}
{Lodato}, G., {Dipierro}, G., {Ragusa}, E., {et~al.} 2019, \mnras, 486, 453

\bibitem[{{Long} {et~al.}(2018){Long}, {Herczeg}, {Pascucci}, {Apai},
  {Henning}, {Manara}, {Mulders}, {Sz{\H{u}}cs}, \& {Hendler}}]{Long2018}
{Long}, F., {Herczeg}, G.~J., {Pascucci}, I., {et~al.} 2018, \apj, 863, 61

\bibitem[{{Long} {et~al.}(2019){Long}, {Herczeg}, {Harsono}, {Pinilla},
  {Tazzari}, {Manara}, {Pascucci}, {Cabrit}, {Nisini}, {Johnstone}, {Edwards},
  {Salyk}, {Menard}, {Lodato}, {Boehler}, {Mace}, {Liu}, {Mulders}, {Hendler},
  {Ragusa}, {Fischer}, {Banzatti}, {Rigliaco}, {van de Plas}, {Dipierro},
  {Gully-Santiago}, \& {Lopez-Valdivia}}]{Long2019}
{Long}, F., {Herczeg}, G.~J., {Harsono}, D., {et~al.} 2019, \apj, 882, 49

\bibitem[{{Long} {et~al.}(2020){Long}, {Pinilla}, {Herczeg}, {Andrews},
  {Harsono}, {Johnstone}, {Ragusa}, {Pascucci}, {Wilner}, {Hendler},
  {Jennings}, {Liu}, {Lodato}, {Menard}, {van de Plas}, \&
  {Dipierro}}]{Long2020}
{Long}, F., {Pinilla}, P., {Herczeg}, G.~J., {et~al.} 2020, \apj, 898, 36

\bibitem[{{Luhman}(2003)}]{Luhman2003}
{Luhman}, K.~L. 2003, in IAU Symposium, Vol. 211, Brown Dwarfs, ed.
  E.~{Mart{\'\i}n}, 103

\bibitem[{{Luhman} {et~al.}(2007){Luhman}, {Allers}, {Jaffe}, {Cushing},
  {Williams}, {Slesnick}, \& {Vacca}}]{Luhman2007}
{Luhman}, K.~L., {Allers}, K.~N., {Jaffe}, D.~T., {et~al.} 2007, \apj, 659,
  1629

\bibitem[{{Maaskant} {et~al.}(2013){Maaskant}, {Honda}, {Waters}, {Tielens},
  {Dominik}, {Min}, {Verhoeff}, {Meeus}, \& {van den Ancker}}]{Maaskant2013}
{Maaskant}, K.~M., {Honda}, M., {Waters}, L.~B.~F.~M., {et~al.} 2013, \aap,
  555, A64

\bibitem[{{Malla} {et~al.}(2020){Malla}, {Stello}, {Huber}, {Montet},
  {Bedding}, {Fredslund Andersen}, {Grundahl}, {Jessen-Hansen}, {Hey}, {Palle},
  {Deng}, {Zhang}, {Chen}, {Lloyd}, \& {Antoci}}]{Malla2020}
{Malla}, S.~P., {Stello}, D., {Huber}, D., {et~al.} 2020, \mnras, 496, 5423

\bibitem[{{Mamajek}(2009)}]{Mamajek2009}
{Mamajek}, E.~E. 2009, in American Institute of Physics Conference Series, Vol.
  1158, American Institute of Physics Conference Series, ed. T.~{Usuda},
  M.~{Tamura}, \& M.~{Ishii}, 3--10

\bibitem[{{Manara} {et~al.}(2014){Manara}, {Testi}, {Natta}, {Rosotti},
  {Benisty}, {Ercolano}, \& {Ricci}}]{Manara2014}
{Manara}, C.~F., {Testi}, L., {Natta}, A., {et~al.} 2014, \aap, 568, A18

\bibitem[{{Manara} {et~al.}(2017){Manara}, {Testi}, {Herczeg}, {Pascucci},
  {Alcal{\'a}}, {Natta}, {Antoniucci}, {Fedele}, {Mulders}, {Henning},
  {Mohanty}, {Prusti}, \& {Rigliaco}}]{Manara2017}
{Manara}, C.~F., {Testi}, L., {Herczeg}, G.~J., {et~al.} 2017, \aap, 604, A127

\bibitem[{{Manara} {et~al.}(2018){Manara}, {Prusti}, {Comeron}, {Mor},
  {Alcal{\'a}}, {Antoja}, {Facchini}, {Fedele}, {Frasca}, {Jerabkova},
  {Rosotti}, {Spezzi}, \& {Spina}}]{Manara2018}
{Manara}, C.~F., {Prusti}, T., {Comeron}, F., {et~al.} 2018, \aap, 615, L1

\bibitem[{{Manara} {et~al.}(2020){Manara}, {Natta}, {Rosotti}, {Alcal{\'a}},
  {Nisini}, {Lodato}, {Testi}, {Pascucci}, {Hillenbrand}, {Carpenter},
  {Scholz}, {Fedele}, {Frasca}, {Mulders}, {Rigliaco}, {Scardoni}, \&
  {Zari}}]{Manara2020}
{Manara}, C.~F., {Natta}, A., {Rosotti}, G.~P., {et~al.} 2020, \aap, 639, A58

\bibitem[{{Matr{\`a}} {et~al.}(2018){Matr{\`a}}, {Marino}, {Kennedy}, {Wyatt},
  {{\"O}berg}, \& {Wilner}}]{Matra2018}
{Matr{\`a}}, L., {Marino}, S., {Kennedy}, G.~M., {et~al.} 2018, \apj, 859, 72

\bibitem[{{Matthews} {et~al.}(2014){Matthews}, {Krivov}, {Wyatt}, {Bryden}, \&
  {Eiroa}}]{Matthews2014}
{Matthews}, B.~C., {Krivov}, A.~V., {Wyatt}, M.~C., {Bryden}, G., \& {Eiroa},
  C. 2014, in Protostars and Planets VI, ed. H.~{Beuther}, R.~S. {Klessen},
  C.~P. {Dullemond}, \& T.~{Henning}, 521

\bibitem[{{Mayor} {et~al.}(2011){Mayor}, {Marmier}, {Lovis}, {Udry},
  {S{\'e}gransan}, {Pepe}, {Benz}, {Bertaux}, {Bouchy}, {Dumusque}, {Lo Curto},
  {Mordasini}, {Queloz}, \& {Santos}}]{Mayor2011}
{Mayor}, M., {Marmier}, M., {Lovis}, C., {et~al.} 2011, arXiv e-prints,
  arXiv:1109.2497

\bibitem[{{McClure} {et~al.}(2010){McClure}, {Furlan}, {Manoj}, {Luhman},
  {Watson}, {Forrest}, {Espaillat}, {Calvet}, {D'Alessio}, {Sargent}, {Tobin},
  \& {Chiang}}]{McClure2010}
{McClure}, M.~K., {Furlan}, E., {Manoj}, P., {et~al.} 2010, \apjs, 188, 75

\bibitem[{{Meeus} {et~al.}(2001){Meeus}, {Waters}, {Bouwman}, {van den Ancker},
  {Waelkens}, \& {Malfait}}]{Meeus2001}
{Meeus}, G., {Waters}, L.~B.~F.~M., {Bouwman}, J., {et~al.} 2001, \aap, 365,
  476

\bibitem[{{Meshkat} {et~al.}(2017){Meshkat}, {Mawet}, {Bryan}, {Hinkley},
  {Bowler}, {Stapelfeldt}, {Batygin}, {Padgett}, {Morales}, {Serabyn},
  {Christiaens}, {Brandt}, \& {Wahhaj}}]{Meshkat2017}
{Meshkat}, T., {Mawet}, D., {Bryan}, M.~L., {et~al.} 2017, \aj, 154, 245

\bibitem[{{Meyer} \& {Wilking}(2009)}]{Meyer2009}
{Meyer}, M.~R., \& {Wilking}, B.~A. 2009, \pasp, 121, 350

\bibitem[{{Michel} {et~al.}(2021){Michel}, {van der Marel}, \&
  {Matthews}}]{Michel2020}
{Michel}, A., {van der Marel}, N., \& {Matthews}, B. 2021, subm. to ApJ

\bibitem[{{Moe} \& {Kratter}(2019)}]{MoeKratter2019}
{Moe}, M., \& {Kratter}, K.~M. 2019, arXiv e-prints, arXiv:1912.01699

\bibitem[{{Mulders}(2018)}]{Mulders2018}
{Mulders}, G.~D. 2018, {Planet Populations as a Function of Stellar
  Properties}, 153

\bibitem[{{Mulders} {et~al.}(2015){Mulders}, {Pascucci}, \&
  {Apai}}]{Mulders2015a}
{Mulders}, G.~D., {Pascucci}, I., \& {Apai}, D. 2015, \apj, 798, 112

\bibitem[{{Mulders} {et~al.}(2018){Mulders}, {Pascucci}, {Apai}, \&
  {Ciesla}}]{Mulders2018epos}
{Mulders}, G.~D., {Pascucci}, I., {Apai}, D., \& {Ciesla}, F.~J. 2018, \aj,
  156, 24

\bibitem[{{Muley} {et~al.}(2019){Muley}, {Fung}, \& {van der
  Marel}}]{Muley2019}
{Muley}, D., {Fung}, J., \& {van der Marel}, N. 2019, \apjl, 879, L2

\bibitem[{{Murphy} {et~al.}(2013){Murphy}, {Lawson}, \& {Bessell}}]{Murphy2013}
{Murphy}, S.~J., {Lawson}, W.~A., \& {Bessell}, M.~S. 2013, \mnras, 435, 1325

\bibitem[{{Najita} \& {Kenyon}(2014)}]{NajitaKenyon2014}
{Najita}, J.~R., \& {Kenyon}, S.~J. 2014, \mnras, 445, 3315

\bibitem[{{Nielsen} {et~al.}(2019){Nielsen}, {De Rosa}, {Macintosh}, {Wang},
  {Ruffio}, {Chiang}, {Marley}, {Saumon}, {Savransky}, {Ammons}, {Bailey},
  {Barman}, {Blain}, {Bulger}, {Burrows}, {Chilcote}, {Cotten}, {Czekala},
  {Doyon}, {Duch{\^e}ne}, {Esposito}, {Fabrycky}, {Fitzgerald}, {Follette},
  {Fortney}, {Gerard}, {Goodsell}, {Graham}, {Greenbaum}, {Hibon}, {Hinkley},
  {Hirsch}, {Hom}, {Hung}, {Dawson}, {Ingraham}, {Kalas}, {Konopacky},
  {Larkin}, {Lee}, {Lin}, {Maire}, {Marchis}, {Marois}, {Metchev},
  {Millar-Blanchaer}, {Morzinski}, {Oppenheimer}, {Palmer}, {Patience},
  {Perrin}, {Poyneer}, {Pueyo}, {Rafikov}, {Rajan}, {Rameau}, {Rantakyr{\"o}},
  {Ren}, {Schneider}, {Sivaramakrishnan}, {Song}, {Soummer}, {Tallis},
  {Thomas}, {Ward-Duong}, \& {Wolff}}]{Nielsen2019}
{Nielsen}, E.~L., {De Rosa}, R.~J., {Macintosh}, B., {et~al.} 2019, \aj, 158,
  13

\bibitem[{{Owen} \& {Clarke}(2012)}]{Owen2012}
{Owen}, J.~E., \& {Clarke}, C.~J. 2012, \mnras, 426, L96

\bibitem[{{Owen} \& {Wu}(2017)}]{OwenWu2017}
{Owen}, J.~E., \& {Wu}, Y. 2017, \apj, 847, 29

\bibitem[{{Paardekooper} {et~al.}(2010){Paardekooper}, {Baruteau}, {Crida}, \&
  {Kley}}]{Paardekooper2010}
{Paardekooper}, S.~J., {Baruteau}, C., {Crida}, A., \& {Kley}, W. 2010, \mnras,
  401, 1950

\bibitem[{{Paardekooper} \& {Papaloizou}(2009)}]{Paardekooper2009}
{Paardekooper}, S.~J., \& {Papaloizou}, J.~C.~B. 2009, \mnras, 394, 2283

\bibitem[{{Pascucci} {et~al.}(2016){Pascucci}, {Testi}, {Herczeg}, {Long},
  {Manara}, {Hendler}, {Mulders}, {Krijt}, {Ciesla}, {Henning}, {Mohanty},
  {Drabek-Maunder}, {Apai}, {Sz{\H{u}}cs}, {Sacco}, \&
  {Olofsson}}]{Pascucci2016}
{Pascucci}, I., {Testi}, L., {Herczeg}, G.~J., {et~al.} 2016, \apj, 831, 125

\bibitem[{{Pecaut} \& {Mamajek}(2016)}]{Pecaut2016}
{Pecaut}, M.~J., \& {Mamajek}, E.~E. 2016, \mnras, 461, 794

\bibitem[{{Pinilla} {et~al.}(2012{\natexlab{a}}){Pinilla}, {Benisty}, \&
  {Birnstiel}}]{Pinilla2012b}
{Pinilla}, P., {Benisty}, M., \& {Birnstiel}, T. 2012{\natexlab{a}}, \aap, 545,
  A81

\bibitem[{{Pinilla} {et~al.}(2012{\natexlab{b}}){Pinilla}, {Birnstiel},
  {Ricci}, {Dullemond}, {Uribe}, {Testi}, \& {Natta}}]{Pinilla2012a}
{Pinilla}, P., {Birnstiel}, T., {Ricci}, L., {et~al.} 2012{\natexlab{b}}, \aap,
  538, A114

\bibitem[{{Pinilla} {et~al.}(2020){Pinilla}, {Pascucci}, \&
  {Marino}}]{Pinilla2020}
{Pinilla}, P., {Pascucci}, I., \& {Marino}, S. 2020, \aap, 635, A105

\bibitem[{{Pinilla} {et~al.}(2018){Pinilla}, {Tazzari}, {Pascucci}, {Youdin},
  {Garufi}, {Manara}, {Testi}, {van der Plas}, {Barenfeld}, {Canovas}, {Cox},
  {Hendler}, {P{\'e}rez}, \& {van der Marel}}]{Pinilla2018}
{Pinilla}, P., {Tazzari}, M., {Pascucci}, I., {et~al.} 2018, \apj, 859, 32

\bibitem[{{Pinte} {et~al.}(2018){Pinte}, {Price}, {M{\'e}nard}, {Duch{\^e}ne},
  {Dent}, {Hill}, {de Gregorio-Monsalvo}, {Hales}, \& {Mentiplay}}]{Pinte2018}
{Pinte}, C., {Price}, D.~J., {M{\'e}nard}, F., {et~al.} 2018, \apjl, 860, L13

\bibitem[{{Pinte} {et~al.}(2019){Pinte}, {van der Plas}, {M{\'e}nard}, {Price},
  {Christiaens}, {Hill}, {Mentiplay}, {Ginski}, {Choquet}, {Boehler},
  {Duch{\^e}ne}, {Perez}, \& {Casassus}}]{Pinte2019}
{Pinte}, C., {van der Plas}, G., {M{\'e}nard}, F., {et~al.} 2019, Nature
  Astronomy, 419

\bibitem[{{Pinte} {et~al.}(2020){Pinte}, {Price}, {M{\'e}nard}, {Duch{\^e}ne},
  {Christiaens}, {Andrews}, {Huang}, {Hill}, {van der Plas}, {Perez}, {Isella},
  {Boehler}, {Dent}, {Mentiplay}, \& {Loomis}}]{Pinte2020}
{Pinte}, C., {Price}, D.~J., {M{\'e}nard}, F., {et~al.} 2020, \apjl, 890, L9

\bibitem[{{Preibisch} \& {Mamajek}(2008)}]{Preibisch2008}
{Preibisch}, T., \& {Mamajek}, E. 2008, {The Nearest OB Association:
  Scorpius-Centaurus (Sco OB2)}, ed. B.~{Reipurth}, Vol.~5, 235

\bibitem[{{Price} {et~al.}(2018){Price}, {Cuello}, {Pinte}, {Mentiplay},
  {Casassus}, {Christiaens}, {Kennedy}, {Cuadra}, {Sebastian Perez}, {Marino},
  {Armitage}, {Zurlo}, {Juhasz}, {Ragusa}, {Laibe}, \& {Lodato}}]{Price2018}
{Price}, D.~J., {Cuello}, N., {Pinte}, C., {et~al.} 2018, \mnras, 477, 1270

\bibitem[{{Rab} {et~al.}(2020){Rab}, {Kamp}, {Dominik}, {Ginski}, {Muro-Arena},
  {Thi}, {Waters}, \& {Woitke}}]{Rab2019}
{Rab}, C., {Kamp}, I., {Dominik}, C., {et~al.} 2020, \aap, 642, A165

\bibitem[{{Raghavan} {et~al.}(2010){Raghavan}, {McAlister}, {Henry}, {Latham},
  {Marcy}, {Mason}, {Gies}, {White}, \& {ten Brummelaar}}]{Raghavan2010}
{Raghavan}, D., {McAlister}, H.~A., {Henry}, T.~J., {et~al.} 2010, \apjs, 190,
  1

\bibitem[{{Reffert} {et~al.}(2015){Reffert}, {Bergmann}, {Quirrenbach},
  {Trifonov}, \& {K{\"u}nstler}}]{Reffert2015}
{Reffert}, S., {Bergmann}, C., {Quirrenbach}, A., {Trifonov}, T., \&
  {K{\"u}nstler}, A. 2015, \aap, 574, A116

\bibitem[{{Ribas} {et~al.}(2015){Ribas}, {Bouy}, \& {Mer{\'{\i}}n}}]{Ribas2015}
{Ribas}, {\'A}., {Bouy}, H., \& {Mer{\'{\i}}n}, B. 2015, \aap, 576, A52

\bibitem[{{Rodriguez} {et~al.}(2015){Rodriguez}, {van der Plas}, {Kastner},
  {Schneider}, {Faherty}, {Mardones}, {Mohanty}, \& {Principe}}]{Rodriguez2015}
{Rodriguez}, D.~R., {van der Plas}, G., {Kastner}, J.~H., {et~al.} 2015, \aap,
  582, L5

\bibitem[{{Rogers} \& {Owen}(2021)}]{RogersOwen2020}
{Rogers}, J.~G., \& {Owen}, J.~E. 2021, \mnras, 503, 1526

\bibitem[{{Rogers}(2015)}]{Rogers2015}
{Rogers}, L.~A. 2015, \apj, 801, 41

\bibitem[{{Rosotti} {et~al.}(2016){Rosotti}, {Juhasz}, {Booth}, \&
  {Clarke}}]{Rosotti2016}
{Rosotti}, G.~P., {Juhasz}, A., {Booth}, R.~A., \& {Clarke}, C.~J. 2016,
  \mnras, 459, 2790

\bibitem[{{Rosotti} {et~al.}(2019){Rosotti}, {Benisty}, {Juh{\'a}sz}, {Teague},
  {Clarke}, {Dominik}, {Dullemond}, {Klaassen}, {Matra}, \&
  {Stolker}}]{Rosotti2019}
{Rosotti}, G.~P., {Benisty}, M., {Juh{\'a}sz}, A., {et~al.} 2019, \mnras, 2689

\bibitem[{{Rugel} {et~al.}(2018){Rugel}, {Fedele}, \& {Herczeg}}]{Rugel2018}
{Rugel}, M., {Fedele}, D., \& {Herczeg}, G. 2018, \aap, 609, A70

\bibitem[{{Ru{\'\i}z-Rodr{\'\i}guez}
  {et~al.}(2018{\natexlab{a}}){Ru{\'\i}z-Rodr{\'\i}guez}, {Cieza}, {Williams},
  {Andrews}, {Principe}, {Caceres}, {Canovas}, {Casassus}, {Schreiber}, \&
  {Kastner}}]{Ruiz-Rodriguez2018}
{Ru{\'\i}z-Rodr{\'\i}guez}, D., {Cieza}, L.~A., {Williams}, J.~P., {et~al.}
  2018{\natexlab{a}}, \mnras, 478, 3674

\bibitem[{{Ru{\'\i}z-Rodr{\'\i}guez}
  {et~al.}(2018{\natexlab{b}}){Ru{\'\i}z-Rodr{\'\i}guez}, {Cieza}, {Williams},
  {Andrews}, {Principe}, {Caceres}, {Canovas}, {Casassus}, {Schreiber}, \&
  {Kastner}}]{RuizRodriguez2018}
---. 2018{\natexlab{b}}, \mnras, 478, 3674

\bibitem[{{Ru{\'\i}z-Rodr{\'\i}guez}
  {et~al.}(2018{\natexlab{c}}){Ru{\'\i}z-Rodr{\'\i}guez}, {Cieza}, {Williams},
  {Andrews}, {Principe}, {Caceres}, {Canovas}, {Casassus}, {Schreiber}, \&
  {Kastner}}]{Ruiz2018}
---. 2018{\natexlab{c}}, \mnras, 478, 3674

\bibitem[{{Santerne} {et~al.}(2016){Santerne}, {Moutou}, {Tsantaki}, {Bouchy},
  {H{\'e}brard}, {Adibekyan}, {Almenara}, {Amard}, {Barros}, {Boisse},
  {Bonomo}, {Bruno}, {Courcol}, {Deleuil}, {Demangeon}, {D{\'\i}az}, {Guillot},
  {Havel}, {Montagnier}, {Rajpurohit}, {Rey}, \& {Santos}}]{Santerne2016}
{Santerne}, A., {Moutou}, C., {Tsantaki}, M., {et~al.} 2016, \aap, 587, A64

\bibitem[{{Sellek} {et~al.}(2020){Sellek}, {Booth}, \& {Clarke}}]{Sellek2020}
{Sellek}, A.~D., {Booth}, R.~A., \& {Clarke}, C.~J. 2020, arXiv e-prints,
  arXiv:2008.07530

\bibitem[{{Sibthorpe} {et~al.}(2018){Sibthorpe}, {Kennedy}, {Wyatt},
  {Lestrade}, {Greaves}, {Matthews}, \& {Duch{\^e}ne}}]{Sibthorpe2018}
{Sibthorpe}, B., {Kennedy}, G.~M., {Wyatt}, M.~C., {et~al.} 2018, \mnras, 475,
  3046

\bibitem[{{Sicilia-Aguilar} {et~al.}(2009){Sicilia-Aguilar}, {Bouwman},
  {Juh{\'a}sz}, {Henning}, {Roccatagliata}, {Lawson}, {Acke}, {Feigelson},
  {Tielens}, {Decin}, \& {Meeus}}]{Sicilia2009}
{Sicilia-Aguilar}, A., {Bouwman}, J., {Juh{\'a}sz}, A., {et~al.} 2009, \apj,
  701, 1188

\bibitem[{{Simon} {et~al.}(2019){Simon}, {Guilloteau}, {Beck}, {Chapillon}, {Di
  Folco}, {Dutrey}, {Feiden}, {Grosso}, {Pi{\'e}tu}, {Prato}, \&
  {Schaefer}}]{Simon2019}
{Simon}, M., {Guilloteau}, S., {Beck}, T.~L., {et~al.} 2019, \apj, 884, 42

\bibitem[{{Sinclair} {et~al.}(2020){Sinclair}, {Rosotti}, {Juhasz}, \&
  {Clarke}}]{Sinclair2020}
{Sinclair}, C.~A., {Rosotti}, G.~P., {Juhasz}, A., \& {Clarke}, C.~J. 2020,
  \mnras, 493, 3535

\bibitem[{{Sokal} {et~al.}(2018){Sokal}, {Deen}, {Mace}, {Lee}, {Oh}, {Kim},
  {Kidder}, \& {Jaffe}}]{Sokal2018}
{Sokal}, K.~R., {Deen}, C.~P., {Mace}, G.~N., {et~al.} 2018, \apj, 853, 120

\bibitem[{{Strom} {et~al.}(1989){Strom}, {Strom}, {Edwards}, {Cabrit}, \&
  {Skrutskie}}]{Strom1989}
{Strom}, K.~M., {Strom}, S.~E., {Edwards}, S., {Cabrit}, S., \& {Skrutskie},
  M.~F. 1989, \aj, 97, 1451

\bibitem[{{Suzuki} {et~al.}(2016){Suzuki}, {Bennett}, {Sumi}, {Bond}, {Rogers},
  {Abe}, {Asakura}, {Bhattacharya}, {Donachie}, {Freeman}, {Fukui}, {Hirao},
  {Itow}, {Koshimoto}, {Li}, {Ling}, {Masuda}, {Matsubara}, {Muraki},
  {Nagakane}, {Onishi}, {Oyokawa}, {Rattenbury}, {Saito}, {Sharan}, {Shibai},
  {Sullivan}, {Tristram}, {Yonehara}, \& {MOA Collaboration}}]{Suzuki2016}
{Suzuki}, D., {Bennett}, D.~P., {Sumi}, T., {et~al.} 2016, \apj, 833, 145

\bibitem[{{Takahashi} \& {Inutsuka}(2014)}]{Takahashi2016}
{Takahashi}, S.~Z., \& {Inutsuka}, S.-i. 2014, \apj, 794, 55

\bibitem[{{Tang} {et~al.}(2017){Tang}, {Guilloteau}, {Dutrey}, {Muto}, {Shen},
  {Gu}, {Inutsuka}, {Momose}, {Pietu}, {Fukagawa}, {Chapillon}, {Ho}, {di
  Folco}, {Corder}, {Ohashi}, \& {Hashimoto}}]{Tang2017}
{Tang}, Y.-W., {Guilloteau}, S., {Dutrey}, A., {et~al.} 2017, \apj, 840, 32

\bibitem[{{Teague} {et~al.}(2019){Teague}, {Bae}, \& {Bergin}}]{Teague2019}
{Teague}, R., {Bae}, J., \& {Bergin}, E.~A. 2019, \nat, 574, 378

\bibitem[{{Teague} {et~al.}(2018){Teague}, {Bae}, {Bergin}, {Birnstiel}, \&
  {Foreman-Mackey}}]{Teague2018}
{Teague}, R., {Bae}, J., {Bergin}, E.~A., {Birnstiel}, T., \& {Foreman-Mackey},
  D. 2018, \apjl, 860, L12

\bibitem[{{Testi} {et~al.}(2014){Testi}, {Birnstiel}, {Ricci}, {Andrews},
  {Blum}, {Carpenter}, {Dominik}, {Isella}, {Natta}, {Williams}, \&
  {Wilner}}]{Testi2014}
{Testi}, L., {Birnstiel}, T., {Ricci}, L., {et~al.} 2014, in Protostars and
  Planets VI, ed. H.~{Beuther}, R.~S. {Klessen}, C.~P. {Dullemond}, \&
  T.~{Henning}, 339

\bibitem[{{Thompson} {et~al.}(2018){Thompson}, {Coughlin}, {Hoffman},
  {Mullally}, {Christiansen}, {Burke}, {Bryson}, {Batalha}, {Haas},
  {Catanzarite}, {Rowe}, {Barentsen}, {Caldwell}, {Clarke}, {Jenkins}, {Li},
  {Latham}, {Lissauer}, {Mathur}, {Morris}, {Seader}, {Smith}, {Klaus},
  {Twicken}, {Van Cleve}, {Wohler}, {Akeson}, {Ciardi}, {Cochran}, {Henze},
  {Howell}, {Huber}, {Pr{\v{s}}a}, {Ram{\'\i}rez}, {Morton}, {Barclay},
  {Campbell}, {Chaplin}, {Charbonneau}, {Christensen-Dalsgaard}, {Dotson},
  {Doyle}, {Dunham}, {Dupree}, {Ford}, {Geary}, {Girouard}, {Isaacson},
  {Kjeldsen}, {Quintana}, {Ragozzine}, {Shabram}, {Shporer}, {Silva Aguirre},
  {Steffen}, {Still}, {Tenenbaum}, {Welsh}, {Wolfgang}, {Zamudio}, {Koch}, \&
  {Borucki}}]{Thompson2018}
{Thompson}, S.~E., {Coughlin}, J.~L., {Hoffman}, K., {et~al.} 2018, \apjs, 235,
  38

\bibitem[{{Thureau} {et~al.}(2014){Thureau}, {Greaves}, {Matthews}, {Kennedy},
  {Phillips}, {Booth}, {Duch{\^e}ne}, {Horner}, {Rodriguez}, {Sibthorpe}, \&
  {Wyatt}}]{Thureau2014}
{Thureau}, N.~D., {Greaves}, J.~S., {Matthews}, B.~C., {et~al.} 2014, \mnras,
  445, 2558

\bibitem[{{Tripathi} {et~al.}(2017){Tripathi}, {Andrews}, {Birnstiel}, \&
  {Wilner}}]{Tripathi2017}
{Tripathi}, A., {Andrews}, S.~M., {Birnstiel}, T., \& {Wilner}, D.~J. 2017,
  \apj, 845, 44

\bibitem[{{Tychoniec} {et~al.}(2020){Tychoniec}, {Manara}, {Rosotti}, {van
  Dishoeck}, {Cridland }, {Hsieh}, {Murillo}, {Segura-Cox}, {van Terwisga}, \&
  {Tobin}}]{Tychoniec2020}
{Tychoniec}, {\L}., {Manara}, C.~F., {Rosotti}, G.~P., {et~al.} 2020, \aap,
  640, A19

\bibitem[{{van der Marel} {et~al.}(2019){van der Marel}, {Dong}, {di
  Francesco}, {Williams}, \& {Tobin}}]{vanderMarel2019}
{van der Marel}, N., {Dong}, R., {di Francesco}, J., {Williams}, J.~P., \&
  {Tobin}, J. 2019, \apj, 872, 112

\bibitem[{{van der Marel} {et~al.}(2016){van der Marel}, {Verhaar}, {van
  Terwisga}, {Mer{\'\i}n}, {Herczeg}, {Ligterink}, \& {van
  Dishoeck}}]{vanderMarel2016}
{van der Marel}, N., {Verhaar}, B.~W., {van Terwisga}, S., {et~al.} 2016, \aap,
  592, A126

\bibitem[{{van der Marel} {et~al.}(2018{\natexlab{a}}){van der Marel},
  {Williams}, \& {Bruderer}}]{vanderMarel2018-hd16}
{van der Marel}, N., {Williams}, J.~P., \& {Bruderer}, S. 2018{\natexlab{a}},
  \apjl, 867, L14

\bibitem[{{van der Marel} {et~al.}(2018{\natexlab{b}}){van der Marel},
  {Williams}, {Ansdell}, {Manara}, {Miotello}, {Tazzari}, {Testi},
  {Hogerheijde}, {Bruderer}, {van Terwisga}, \& {van
  Dishoeck}}]{vanderMarel2018}
{van der Marel}, N., {Williams}, J.~P., {Ansdell}, M., {et~al.}
  2018{\natexlab{b}}, \apj, 854, 177

\bibitem[{{van der Marel} {et~al.}(2021){van der Marel}, {Birnstiel}, {Garufi},
  {Ragusa}, {Christiaens}, {Price}, {Sallum}, {Muley}, {Francis}, \&
  {Dong}}]{vanderMarel2020}
{van der Marel}, N., {Birnstiel}, T., {Garufi}, A., {et~al.} 2021, \aj, 161, 33

\bibitem[{{van der Plas} {et~al.}(2016){van der Plas}, {M{\'e}nard},
  {Ward-Duong}, {Bulger}, {Harvey}, {Pinte}, {Patience}, {Hales}, \&
  {Casassus}}]{vanderPlas2016}
{van der Plas}, G., {M{\'e}nard}, F., {Ward-Duong}, K., {et~al.} 2016, \apj,
  819, 102

\bibitem[{{van Terwisga} {et~al.}(2018){van Terwisga}, {van Dishoeck},
  {Ansdell}, {van der Marel}, {Testi}, {Williams}, {Facchini}, {Tazzari},
  {Hogerheijde}, {Trapman}, {Manara}, {Miotello}, {Maud}, \&
  {Harsono}}]{vanTerwisga2018}
{van Terwisga}, S.~E., {van Dishoeck}, E.~F., {Ansdell}, M., {et~al.} 2018,
  \aap, 616, A88

\bibitem[{{Venuti} {et~al.}(2019){Venuti}, {Stelzer}, {Alcal{\'a}}, {Manara},
  {Frasca}, {Jayawardhana}, {Antoniucci}, {Argiroffi}, {Natta}, {Nisini},
  {Randich}, \& {Scholz}}]{Venuti2019}
{Venuti}, L., {Stelzer}, B., {Alcal{\'a}}, J.~M., {et~al.} 2019, \aap, 632, A46

\bibitem[{{Villenave} {et~al.}(2019){Villenave}, {Benisty}, {Dent},
  {M{\'e}nard}, {Garufi}, {Ginski}, {Pinilla}, {Pinte}, {Williams}, {de Boer},
  {Morino}, {Fukagawa}, {Dominik}, {Flock}, {Henning}, {Juh{\'a}sz}, {Keppler},
  {Muro-Arena}, {Olofsson}, {P{\'e}rez}, {van der Plas}, {Zurlo}, {Carle},
  {Feautrier}, {Pavlov}, {Pragt}, {Ramos}, {Sauvage}, {Stadler}, \&
  {Weber}}]{Villenave2019}
{Villenave}, M., {Benisty}, M., {Dent}, W.~R.~F., {et~al.} 2019, \aap, 624, A7

\bibitem[{{Vioque} {et~al.}(2018){Vioque}, {Oudmaijer}, {Baines},
  {Mendigut{\'\i}a}, \& {P{\'e}rez-Mart{\'\i}nez}}]{Vioque2018}
{Vioque}, M., {Oudmaijer}, R.~D., {Baines}, D., {Mendigut{\'\i}a}, I., \&
  {P{\'e}rez-Mart{\'\i}nez}, R. 2018, \aap, 620, A128

\bibitem[{{Wagner} {et~al.}(2019){Wagner}, {Apai}, \& {Kratter}}]{Wagner2019}
{Wagner}, K., {Apai}, D., \& {Kratter}, K.~M. 2019, \apj, 877, 46

\bibitem[{{Walsh} {et~al.}(2011){Walsh}, {Morbidelli}, {Raymond}, {O'Brien}, \&
  {Mandell}}]{Walsh2011}
{Walsh}, K.~J., {Morbidelli}, A., {Raymond}, S.~N., {O'Brien}, D.~P., \&
  {Mandell}, A.~M. 2011, \nat, 475, 206

\bibitem[{{Ward-Duong} {et~al.}(2018){Ward-Duong}, {Patience}, {Bulger}, {van
  der Plas}, {M{\'e}nard}, {Pinte}, {Jackson}, {Bryden}, {Turner}, {Harvey},
  {Hales}, \& {De Rosa}}]{WardDuong2018}
{Ward-Duong}, K., {Patience}, J., {Bulger}, J., {et~al.} 2018, \aj, 155, 54

\bibitem[{{Weidenschilling}(1977)}]{Weidenschilling1977}
{Weidenschilling}, S.~J. 1977, \mnras, 180, 57

\bibitem[{{Wichittanakom} {et~al.}(2020){Wichittanakom}, {Oudmaijer},
  {Fairlamb}, {Mendigut{\'\i}a}, {Vioque}, \& {Ababakr}}]{Wichittanakom2020}
{Wichittanakom}, C., {Oudmaijer}, R.~D., {Fairlamb}, J.~R., {et~al.} 2020,
  \mnras, 493, 234

\bibitem[{{Wilking} {et~al.}(2008){Wilking}, {Gagn{\'e}}, \&
  {Allen}}]{Wilking2008}
{Wilking}, B.~A., {Gagn{\'e}}, M., \& {Allen}, L.~E. 2008, {Star Formation in
  the {\ensuremath{\rho}} Ophiuchi Molecular Cloud}, ed. B.~{Reipurth}, Vol.~5,
  351

\bibitem[{{Winter} {et~al.}(2019){Winter}, {Clarke}, {Rosotti}, {Hacar}, \&
  {Alexander}}]{Winter2019}
{Winter}, A.~J., {Clarke}, C.~J., {Rosotti}, G.~P., {Hacar}, A., \&
  {Alexander}, R. 2019, \mnras, 490, 5478

\bibitem[{{Wyatt}(2008)}]{Wyatt2008}
{Wyatt}, M.~C. 2008, \araa, 46, 339

\bibitem[{{Yang} {et~al.}(2020){Yang}, {Xie}, \& {Zhou}}]{Yang2020}
{Yang}, J.-Y., {Xie}, J.-W., \& {Zhou}, J.-L. 2020, \aj, 159, 164

\bibitem[{{Yelverton} {et~al.}(2020){Yelverton}, {Kennedy}, \&
  {Su}}]{Yelverton2020}
{Yelverton}, B., {Kennedy}, G.~M., \& {Su}, K. Y.~L. 2020, \mnras, 495, 1943

\bibitem[{{Zhang} {et~al.}(2015){Zhang}, {Blake}, \& {Bergin}}]{Zhang2015}
{Zhang}, K., {Blake}, G.~A., \& {Bergin}, E.~A. 2015, \apjl, 806, L7

\bibitem[{{Zhang} {et~al.}(2018){Zhang}, {Liu}, {Best}, {Magnier}, {Aller},
  {Chambers}, {Draper}, {Flewelling}, {Hodapp}, {Kaiser}, {Kudritzki},
  {Metcalfe}, {Wainscoat}, \& {Waters}}]{Zhang2018}
{Zhang}, Z., {Liu}, M.~C., {Best}, W. M.~J., {et~al.} 2018, \apj, 858, 41

\bibitem[{{Zhu} {et~al.}(2018){Zhu}, {Petrovich}, {Wu}, {Dong}, \&
  {Xie}}]{Zhu2018}
{Zhu}, W., {Petrovich}, C., {Wu}, Y., {Dong}, S., \& {Xie}, J. 2018, \apj, 860,
  101

\end{thebibliography}

\appendix

\begin{table}[!ht]
\centering
\caption{Full sample}
\label{tbl:fullsample}
\begin{tabular}{llllllllllll}
\hline
Region&Target&$d$&$F_{\rm cont}$&Band&$M_{\rm dust}$&$R_{\rm size}$&SpT&$M_*$&Class\footnote{T = transition disk, R = ring disk, E = extended disk ($>$40 au), C = compact, N = non-detection}&Ref. ALMA & Ref. stellar \\
&&(pc)&(mJy)&&($M_{\oplus}$)&(au)&&($M_{\odot}$)&&& \\
\hline
Chamaeleon&	J11080329-7739174&	185&	2253 &7&322	&	179	&A0&	2.40&T&	1 & 32 \\
Chamaeleon&	J10533978-7712338&	192&	4.6&7&0.71	&$<$	12&	M2&	0.33 &C&2& 33 \\
Chamaeleon&	J10555973-7724399&	185&	34.1&7&4.91	&	17&	K7&	0.81	&C&2 & 33 \\
Chamaeleon&	J10563044-7711393&	183&	117.6&7&16.5	&	137&	K7&	0.76	&T&2 & 33 \\
Chamaeleon&	J10574219-7659356&	185$^*$&	9.12&7&1.30	&	$<$	11&	M3&	0.32	&C&2 & 33 \\
\hline
\end{tabular}
\\
$^*$) No parallax measurement from Gaia DR2, so the average distance to the cloud is assumed.\\
Refs. 1) \citet{vanderPlas2016},
2) \citet{Pascucci2016},
3) \citet{Cazzoletti2019},
4) \citet{Francis2020},
5) \citet{Villenave2019},
6) Aguayo et al. in prep.,
7) \citet{Ruiz2018},
8) \citet{Andrews2018},
9) \citet{Ansdell2018},
10) \citet{vanTerwisga2018},
11) \citet{Cieza2019},
12) \citet{Andrews2013},
13) \citet{Tripathi2017},
14) \citet{Bacciotti2018},
15) \citet{Akeson2014},
16) \citet{WardDuong2018},
17) \citet{Facchini2019},
18) \citet{Akeson2019},
19) \citet{Long2019},
20) \citet{Hardy2015},
21) \citet{Simon2019},
22) \citet{Andrews2016},
23) \citet{Tang2017},
24) \citet{Andrews2009},
25) \citet{Pinilla2018},
26) \citet{Rodriguez2015},
27) \citet{vanderMarel2016},
28) \citet{Ansdell2020},
29) \citet{Casassus2013},
30) \citet{Keppler2019},
31) \citet{Barenfeld2016},

32) \citet{Wichittanakom2020},
33) \citet{Manara2018},
34) \citet{Manara2017},
35) \citet{Manara2014},
36) \citet{Murphy2013},
37) \citet{Vioque2018},
38) \citet{Rugel2018},
39) \citet{Alcala2017},
40) \citet{Sokal2018},
41) \citet{Venuti2019},
42) \citet{Manara2020},
43) \citet{McClure2010}
\\
This table is available in online format only. The first 5 entries are shown for guidance.
\end{table}

\end{document}